\documentstyle[11pt,amssymb]{article}

\makeatletter
\@addtoreset{equation}{section}
\makeatother

\def\dalemb#1#2{{\vbox{\hrule height .#2pt
        \hbox{\vrule width.#2pt height#1pt \kern#1pt
                \vrule width.#2pt}
        \hrule height.#2pt}}}

\def\0{{\sst{(0)}}}
\def\1{{\sst{(1)}}}
\def\2{{\sst{(2)}}}
\def\3{{\sst{(3)}}}
\def\4{{\sst{(4)}}}
\def\5{{\sst{(5)}}}
\def\6{{\sst{(6)}}}
\def\7{{\sst{(7)}}}
\def\8{{\sst{(8)}}}

\def\ep{\epsilon}
\def\td{\tilde}

\def\half{{\textstyle{1\over2}}}

\def\qu{{\textstyle{1\over 4}}}

\let\a=\alpha \let\b=\beta \let\g=\gamma \let\d=\delta 
    \let\k=\kappa
\let\l=\lambda    
\let\s=\sigma \let\t=\tau    
\let\w=\omega    
    
    \let\G=\Gamma
\let\la=\label  
  
\def\nn{\nonumber} \def\bd{\begin{document}} \def\ed{\end{document}}
\def\ds{\documentstyle} \let\fr=\frac \let\bl=\bigl \let\br=\bigr
\let\Br=\Bigr \let\Bl=\Bigl
\let\bm=\bibitem
\let\na=\nabla
\let\pa=\partial \let\ov=\overline
\newcommand{\be}{\begin{equation}}
\newcommand{\ee}{\end{equation}}
\def\ba{\begin{array}}
\def\ea{\end{array}}
\def\ft#1#2{{\textstyle{{\scriptstyle #1}\over {\scriptstyle #2}}}}
\def\fft#1#2{{#1 \over #2}}
\def\del{\partial}
\def\sst#1{{\scriptscriptstyle #1}}
 \def\oneone{\rlap 1\mkern4mu{\rm l}}
\def\ie{{\it i.e.\ }}
\def\via{{\it via}}
\def\semi{{\ltimes}}
\def\str{{\rm str}}
\def\Dm{{{D_{\sst{max}}}}}
\def\vac{ \left | 0; k \right \rangle }
\def\kvac{ \left | k \right \rangle }

\def\sp{\; \;}

\def\bol{ \left | \phi \right \rangle}
\def\bo1{ \left | B^0 (p^+) \right \rangle}

\def\bolt{ \left | B (p^+) \right \rangle_{\t}}

\def\boxl{ \left | B (x^-) \right \rangle}

\def\barray{ \left | B_{\rm{array}} \right \rangle}

\newcommand{\hsp}{\hspace{0.5cm}}

\newcommand{\ho}[1]{$\, ^{#1}$}
\newcommand{\hoch}[1]{$\, ^{#1}$}
\newcommand{\bea}{\begin{eqnarray}}
\newcommand{\eea}{\end{eqnarray}}
\newcommand{\ra}{\rightarrow}
\newcommand{\lra}{\longrightarrow}
\newcommand{\Lra}{\Leftrightarrow}
\newcommand{\ap}{\alpha^\prime}
\newcommand{\bp}{\tilde \beta^\prime}
\newcommand{\tr}{{\rm tr} }
\newcommand{\Tr}{{\rm Tr} }
\newcommand{\NP}{Nucl. Phys. }

\newcommand{\ams}{{\it Institute for Theoretical Physics,
University of Amsterdam, \\
Valckenierstraat 65, 1018XE Amsterdam, The Netherlands} \\
{\tt email:taylor@science.uva.nl}}
\newcommand{\auth}{{Marika Taylor\hoch{\dagger}}}

\begin{document}
\begin{flushright}
\hfill{\bf hep-th/0507223}\\
\hfill{ITFA-2005-24}
\end{flushright}

\vspace{15pt}

\begin{center}

{\Large \bf General 2 charge geometries}

\vspace{20pt}

\auth
\vspace{15pt}

{\hoch\dagger \ams}

\vspace{15pt}

\underline{ABSTRACT}
\end{center}

Two charge BPS horizon free supergravity geometries are important in
proposals for understanding black hole microstates. In this paper we
construct a new class of geometries in the NS1-P system, corresponding
to solitonic strings carrying fermionic as well as bosonic
condensates. Such geometries are required to account for the full
microscopic entropy of the NS1-P system. We then briefly discuss the
properties of the corresponding geometries in the dual D1-D5 system.

\noindent

\pagebreak
\setcounter{page}{1}

\tableofcontents
\addtocontents{toc}{\protect\setcounter{tocdepth}{1}}

\section{Introduction}

In recent years there has been considerable progress in the
classification of supersymmetric solutions of supergravity
theories. For example, the work of \cite{Gauntlett:2002nw} constructed the most
general supersymmetric solutions of minimal five-dimensional
supergravity which have a timelike killing spinor, and there have
been many subsequent papers extending such analysis to other
supergravity theories.

In the context of string theory, however, this is only part of
what is required to classify the supersymmetric backgrounds of
interest. In string theory, one must also allow for source terms
in the supergravity equations of motion which arise from allowed
objects in the string theory, namely branes and solitonic fundamental
strings. The question of
classifying supergravity backgrounds therefore also implicitly involves a
classification of allowed sources.

This issue was appreciated even before the advent of D-branes. For
example, \cite{Dabholkar:1989jt,Dabholkar:1990yf, Dabholkar:1995nc, Sen, Callan:1995hn}
discussed BPS supergravity solutions corresponding to certain BPS
fundamental string states; the supergravity fields have
sources which are exactly those needed to match the solitonic
fundamental string. After the discovery of D-branes, this
correspondence became frequently used: one associates the boundary
state or open string description of the D-brane with a
corresponding supergravity solution which generically has sources.
(In certain cases, such as the D3-brane and the D1-D5 brane system, the source is actually
not present, but generically for branes there will be source terms
in the supergravity equations of motion. Indeed, generically a system such as the D1-D5 brane system
in which there are no sources can be related by duality to another system (such as the F1-P)
in which there are sources.) This relationship of
course underpins the AdS/CFT conjecture
\cite{Maldacena:1997re,Gubser:1998bc,Witten:1998qj} where on one
side of the duality one takes a decoupling limit of the
supergravity background sourced by the D-branes and on the other
side a corresponding limit of the open string description of the
branes.

Usually in directly constructing solitonic string or brane
backgrounds, one considers only supergravity solutions
corresponding to solitons with bosonic excitations. For example,
\cite{Dabholkar:1995nc} discussed BPS fundamental strings with no
fermionic excitations which carried a null wave in the transverse
plane. Note that in the context of AdS/CFT one is
interested in switching on supersymmetry preserving vacuum
expectation values or operator deformations. Thus implicitly one
is discussing solutions in which one has switched on fermionic
excitations on the branes.

Generically, switching on fermion excitations on branes or strings
gives rise to supergravity backgrounds which involve harmonic
functions with different harmonics than on the static brane,
That is, the sources in the
supergravity equations of motion have a different structure than
when the excitations are purely bosonic. We will discuss this in
detail later in the paper but there are two intuitive ways to
understand this. The first is to compare the couplings between the
supergravity metric (and other fields) and the bosonic and
fermionic excitations on the brane. Take the case of most interest in
this paper, the fundamental string.
Then the bosonic excitations $X^I$ couple to the metric $g_{IJ}$, schematically as
\be
\int d^2 x g_{IJ} \pa X^{I} \pa X^{J},
\ee
but the fermions $\psi^I$ couple not just to the metric but also to
the connection $\G_{IJK}$ via
\be
\int d^2 x \G_{IJK} \psi^{I} \pa X^J \psi^K.
\ee
(The exact action in a general NS-NS background involves also
couplings to the Riemann tensor, and will be given later,
but this does not affect the general argument given here.) Since the
fermions couple to the derivative of the metric,
rather than the metric itself, one would expect that fermionic
condensates give rise to subleading behavior in the
harmonic functions they source. Indeed this turns out to be the case,
as we will show in detail here: the bosonic terms
give rise to delta function source terms, whereas the fermions lead to
$l=1$ source terms, of the form $T^{i} \pa_{i} \d(x)$ where $T^i$ is
some given vector determined by the worldsheet fermion sources.

A second and rather more generic way to see that fermionic condensates
lead to different harmonics in the supergravity
fields is via the AdS/CFT correspondence. Operators built from
fermionic bilinears generally have different dimensions and
R-charges to those of purely bosonic operators. The usual AdS/CFT
dictionary tells us that the asymptotic behavior of fields
in the dual supergravity background is determined via the dimensions
and R-charges of these operators. Thus the (subleading) supergravity
asymptotics of solutions dual to theories in which there are vacuum
expectation values (vevs) or deformations of operators in the same multiplet need not
be the same. Typically, a supergravity solution corresponding to a
fermion bilinear vev or deformation may have dipole
moments absent in the purely bosonic case.

This brings us to the main motivation of this paper. Mathur
and collaborators have conjectured
(in a series of papers, see \cite{Mathur:2005zp,Lunin:2001jy,Lunin:2001fv,Lunin:2002bj})
that if one considers the D1-D5 system
(and its generalizations) the corresponding supergravity geometry is
not the naive geometry with near horizon limit
$AdS_3 \times S^3$. The claim is that for each vacuum in the D1-D5 CFT
there is a corresponding supergravity geometry
which has no horizon. Furthermore, for all of the examples known
in the 2-charge system the geometry is actually
non-singular \cite{Maldacena:2000dr,Lunin:2002iz},
a somewhat surprising result since the generic geometry
is certainly not weakly curved everywhere and one would not
expect string corrections on the geometry to be small.

There are a number of arguments in favor of the matching between the
individual geometries and corresponding vacua in
the CFT, such as scattering calculations \cite{Mathur:2005zp}.
Perhaps the clearest way to match the geometries would be to take appropriate
decoupling limits to extract the asymptotically AdS regions, and then
match the near boundary asymptotics to the field theory
via the standard AdS/CFT dictionary. (In particular, one should use
the work of \cite{deBoer:1998ip} and \cite{Maldacena:2000hw}.)
This has not yet been done in
detail, although one does see in the known geometries,
for example, a manifest matching of R-charges (in the cases where part
of the $SO(4)$ R-symmetry is preserved). The new claim
made here is that if one did try to carry out this matching with the
known 2-charge D1-D5 geometries, one would find that these
were insufficient to match with operators in the CFT built from
fermion bilinears. In other words, we would claim that the
most general 2-charge geometry is not known, and is not within the
class of solutions written down previously.

One can see this as follows. The D1-D5 supergravity solutions were
obtained by a series of dualities from the
F1-P system. Now in the latter system the microscopic counting of
BPS states with a given winding number $n_1$ and
momentum $n_p$ around a circle in the geometry is well-known and
rather simple. In the large charge limit, the number of
states behaves as
\be
{\cal N} \sim e^{2 \pi \sqrt{\frac{c n_1 n_p}{6}}},
\ee
where $c$ is the central charge. As we will review in the next section
this formula arises from the number of ways of partitioning
the excitation amongst distinct oscillators in the large charge
limit. To match the counting with the D1-D5 system, one needs to
effectively freeze excitations
in four transverse directions, and thus include only four bosons and
four fermions, giving $c=6$; one does indeed get
agreement with the D1-D5 counting in this way. In previous literature,
the supergravity geometries corresponding to purely bosonic
states were constructed, and it is these that were dualised to give the
geometries in the D1-D5 system. However, we can see that
there are also an exponentially large number of BPS states involving
fermion bilinears, for which the geometries are unknown,
even in the F1-P system. Put another way, we have only four chiral bosons worth of geometries,
whereas to match with the microscopic counting one would expect an additional four chiral fermions worth
of geometries, and these are missing. To get the most general supergravity geometry
within this F1-P two charge system, one needs to know
these geometries and this is the aim of the current paper. The
resulting geometries exemplify the previous
discussions, in that the harmonic functions involve different harmonics
to the purely bosonic geometries.

The plan of the paper is as follows. In \S \ref{ch2} we discuss the quantum
states in the NS1-P system, demonstrating that states with fixed winding and
momentum generically involve fermionic excitations. In \S \ref{ch3} we review
previous constructions of the solitonic string supergravity geometries corresponding
to the quantum states of the NS1-P system which do not involve fermionic excitations.
In \S\ref{ch4},\S\ref{ch5} and \S\ref{ch6} we construct geometries corresponding to the generic
microstate of the NS1-P system. In \S \ref{ch7} we discuss the matching of the geometries
with microstates; the applicability of the supergravity approximation and the corresponding
geometries in the dual D1-D5 system. Finally, in \S \ref{conc} we summarize our results,
and discuss outstanding issues along with directions for future research. Conventions and a number
of technical issues are the subjects of several appendices.

\section{The NS1-P system} \la{ch2}

We consider a two charge system, the NS1-P system, in which the
fundamental string carries winding number $n_1$ around a circle,
momentum $n_p$ along the circle and left moving excitations at
some level $N_L$ but is in the right moving vacuum $N_R =
\frac{1}{2}$. The fundamental string state then manifestly
preserves the supersymmetries originating from the right moving
sector.

The mass of a string state in the NS-NS sector is given by
\be
m^2
= (2 \pi R n_1 T - \frac{n_p}{R})^2 + 8 \pi T (N_L  - \frac{1}{2})
= (2 \pi R n_1 T + \frac{n_p}{R})^2 + 8 \pi T (N_R - \frac{1}{2}),
\ee
where $R$ is the radius of the circle and $T = 1/(2 \pi \a')$
is the tension of the string. In the right moving vacuum
\be
(N_{L} - \frac{1}{2}) = n_1  n_p,
\ee and
\be m = (2 \pi n_1 R T +
\frac{n_p}{R}).
\ee
The level is written in the usual way in terms
of left moving oscillators
\be
N_{L} = \frac{1}{\a'} \sum_{n}
\a_{-n}^{I} \a_{n I} + \frac{2}{\a'} \sum_{r} r \b^{I}_{-r} \b_{Ir},
\ee
with the commutation relations being
\be
\left
[\a_{m}^{I},\a_{n}^{J} \right ] = m \a' \eta^{IJ} \d_{m+n}, \hsp
\left \{\b^{I}_{r},\b^{J}_{s} \right \} = \half \a' \eta^{IJ} \d_{r+s},
\ee
where $r$ is half integral since the worldsheet fermions are
anti-periodic in the NS sector\footnote{Note the unusual conventions
  for the fermionic oscillators, $\b^I_{r}$ rather than the usual
  $b^I_r$. To avoid confusion we reserve $b^{I}_r$ to denote classical
  coefficients of the fermion field mode expansion. Correspondingly we use
  $\a^I_n$ to denote quantum oscillators and $a^I_n$ to denote the coefficients
  of the boson field mode expansions.}.
For $N_L$ macroscopically large
the number of states behaves as \be {\cal N} \sim e^{2 \pi \sqrt{c
N_{L}/6}}, \ee where $c$ is the central charge. Later we will be
interested in making contact
with the D1-D5 system, for which one compactifies four directions on
a torus and freezes all excitations along these directions. This
means that one would only count states involving bosonic and
fermionic oscillators not along these directions, giving an
effective central charge of 6, from the 4 bosonic oscillators and
the 4 fermionic oscillators.

The physical states at this level follow from the GSO projections
along with the usual physical state
conditions $L_{n} \bol = G_{r} \bol = 0$  (where the right moving part
of the state is suppressed since in all that follows it is the right
moving vacuum). The relevant GSO condition is $(-)^{F_L} = 1$ and thus
all states involve an odd number of fermions, i.e. they are built by
acting with bosonic oscillators and even numbers of fermionic
oscillators on the left moving NS vacuum
\be
\left | e; k \right >_{NS} = e_{I} \b^{I}_{-1/2} \vac_{NS},
\ee
for which $k^2 = 0$ and $ e^{I} k_{I} = 0$.

\bigskip

Since each such state is BPS, there should be a corresponding object as
the coupling $\a'$ is increased and the backreaction is taken into account.
First we should clarify exactly what we mean by correspondence. We
will relate BPS states in the closed string spectrum to supergravity
geometries which preserve the same supersymmetries, in a discussion
analogous to that relating D-brane boundary states (again in the
closed string Hilbert space) to supergravity geometries. In
the latter discussions it is frequently implied that the string state and
supergravity background are the same object, at weak and strong
coupling respectively. However this statement is manifestly not well
defined, since a state is not an observable. What one is actually
referring to is a correspondence between non-renormalized observables
in the two limits. Similarly, whilst one may loosely talk about
a BPS asymptotically AdS geometry being dual to a specific BPS
operator in the dual CFT, this is also not a well defined statement,
and should instead be phrased in terms of a correspondence between
observables in the bulk and those in the CFT defined in terms of vevs or
deformations by this operator. We will return to this issue later, but
having clarified what we mean by correspondence let us now turn to the
supergravity solitons.

For BPS states in the fundamental string spectrum, corresponding solitons in supergravity
can be identified in the following way. One adds to the supergravity
action
\be
S = \frac{1}{2 \k^2} \int d^{10}x \sqrt{-g} e^{-2 \phi}
(R + 4 (\partial \phi)^2 - \frac{1}{12} H^2),
\ee
(where RR fields
are suppressed since we consider only NS-NS backgrounds) source
terms for the NS-NS fields arising from the presence of a
macroscopic string. These follow from the sigma model action for a
general NS-NS background, namely \cite{Bergshoeff:1985qr, Das:1989wf, Das:1988zk}
\bea  \la{si}
S_{\s} &=&
\frac{1}{2 \pi \a'} \int d^2\s \left ( 2 (g_{IJ} + B_{IJ})
\pa_{+}X^{I} \pa_{-} X^{J} \right . \\
&& \left . - i g_{IJ} (\psi^{I}_- D_{+}^{(-)} \psi^{J}_- +
   {\psi}^{I}_+ D_{-}^{(+)}
{\psi}^J_+) - \qu R^{(-)}_{IJKL} \psi^{I}_- \psi^{J}_- {\psi}^K_+
{\psi}^L_+  \right ). \nn
\eea
Here the worldsheet metric has been fixed as $\g_{+-} = - \half$ with the
worldsheet gravitino set to zero; the gauge fixing and conventions are
discussed in appendix \ref{apa}. $\psi_{\pm}$ are negative and
positive chirality worldsheet spinors on which the
covariant derivatives act as
\bea
D_{+}^{(-)} \psi^{I}_- &=& ( \pa_{+} \psi^I_- + (\G^{I}_{\sp JK} - \half
H^{I}_{\sp JK}) \pa_{+}X^{J} \psi^K_-); \\
D_{-}^{(+)} {\psi}^I_+ &=& (\pa_{-} {\psi}^{I}_+ + (\G^{I}_{\sp JK} + \half
H^{I}_{\sp JK}) \pa_{-} X^{J} {\psi}^K_+), \nn
\eea
where the connection and torsion are defined in the usual way as
\bea
\G_{IJK} &=& \half (\pa_{K} g_{IJ} + \pa_{J} g_{IK} - \pa_{I} g_{JK}); \\
H_{IJK}  &=& \pa_{I} B_{JK} + \pa_{J} B_{KI} + \pa_{K} B_{IJ}, \nn
\eea
and the curvature includes the torsion, namely
\be
R^{(-)}_{IJKL} = R_{IJKL} + \half (D_{K} H_{LIJ} - D_{L} H_{KIJ})
+ \qu G^{MN} (H_{IKM} H_{JLN} - H_{ILM} H_{JKN}),
\ee
with $R_{IJKL}$ the Riemann tensor. $R^{(-)}_{IJKL}$ is the curvature
of the torsionful connection $\G^{I (-)}_{\sp JK}$, and is related to
that of $\G^{I (+)}_{\sp JK}$ by $R^{(-)}_{IJKL} = R^{(+)}_{KLIJ}$.
In this gauge the superconformal generators are
\bea
T_{++} &=& \frac{1}{4 \pi \a'} (- g_{IJ} \pa_{+} X^{I} \pa_{+} X^J +
\half i
g_{IJ} \psi_{+}^I D^{(+)}_+ \psi_{+}^J); \la{scg} \\
T_{--} &=&  \frac{1}{4 \pi \a'} (- g_{IJ} \pa_{-} X^{I} \pa_{-} X^J +
\half i
g_{IJ} \psi_{-}^I D^{(-)}_-\psi_{-}^J); \nn \\
J_{+} &=& \frac{1}{2\pi \a'}
(g_{IJ} \psi_{+}^I \pa_{+}X^J + \frac{i}{12} H_{IJK}
\psi_{+}^I \psi_{+}^J \psi_{+}^K); \nn \\
J_{-} &=& \frac{1}{2 \pi \a'}
(g_{IJ} \psi_{-}^I \pa_{-} X^J - \frac{i}{12} H_{IJK}
\psi_{-}^{I} \psi_{-}^J \psi_{-}^K), \nn
\eea
where the covariant derivatives act as
\bea
D_{+}^{(+)} \psi_{+}^I &=& (\pa_{+} \psi^I_+ + (\G^{I}_{\sp JK} + \half
H^{I}_{\sp JK}) \pa_{+} X^{J} \psi_{+}^K); \\
D_{-}^{(-)} \psi_{-}^I &=& (\pa_{-} \psi^{I}_- + (\G^{I}_{\sp JK} -
\half H^{I}_{\sp JK}) \pa_{-} X^{J} \psi_{-}^K), \nn
\eea
and for brevity we give the constraints onshell with respect to the
fermion field equations, which are
\be \la{f1}
D_{-}^{(+)}  \psi_{+}^{I} = \frac{i}{4} R^{I}_{\sp JKL} \psi_{+}^J
\psi_{-}^K \psi_{-}^L; \hsp
D_{+}^{(-)} \psi_{-}^{I} = \frac{i}{4} R^{I}_{\sp JKL} \psi_{-}^J
\psi_{+}^K \psi_{+}^L.
\ee
Note that the bosonic field equations are
\bea \la{b1}
D_{+}^{(-)} \pa_{-} X^{I} &\equiv &  (\pa_{+} \pa_{-} X^{I} + (\G^{I}_{\sp
  JK} - \half H^{I}_{\sp JK}) \pa_{+}X^{J} \pa_{-} X^{K}) \\
&=& \frac{i}{2} g^{IM} (\pa_{+} X^L)
\left ( R^{(-)}_{JKLM} \psi_{-}^J \psi_{-}^K
+ R^{(+)}_{JKLM} \psi_{+}^J \psi_{+}^K \right ) +
\frac{1}{8} \pa^{I} R^{(-)}_{JKLM} \psi_{-}^J \psi^{K}_- \psi^L_{+}
\psi_{+}^M, \nn
\eea
where the fermion field equations have again been used.

\bigskip

The aim is then to associate with each BPS state $\bol$ in the NS-NS
sector in flat
space (with giving winding and momentum charges)
a corresponding solution $(X(\s),\psi_-(\s), {\psi}_+(\s))$
of the classical sigma model equations (imposing super-Virasoro
constraints) within the curved background
$(g_{IJ},B_{IJ},\phi)$ which these worldsheet fields source.
In practice this is rather subtle, not least because of the coupling
between the worldsheet and spacetime equations of motion and previous
discussions have been restricted to the bosonic sector \cite{Dabholkar:1990yf, Dabholkar:1995nc}.
That is, only the supergravity solutions corresponding to states
\be \la{st}
\prod_{l} (\a^{I_l}_{-m_l})^{n_l} \left | e; k; n_1; n_p \right >_{NS}
\ee
with total excitation number $N_L$ were discussed. As we will
review in the analysis below, these states correspond to supergravity
solutions describing oscillating
strings carrying a left moving wave whose profile relates to the
specific distribution of the oscillators in (\ref{st}).

More generally, however, a BPS state at level $N_L$ will involve even
numbers of left moving fermionic excitations - and one would like
to know the supergravity background associated with a string carrying
such a fermionic condensate. In order to consistently solve the
coupled worldsheet and supergravity equations, one needs an ansatz
for both the worldsheet and supergravity fields. The former follows
from considering the classical $\a' \rightarrow 0$ limit at zero
string coupling. It is clear from considering the mode expansions of
the worldsheet fields in the flat background that such states can be described by
fields $(X(\s), \psi_-(x^-))$ with ${\psi}_+ = 0$. $\psi_-$ is
purely left moving because of the field equations in flat space;
the fields must
also satisfy super-Virasoro constraints and give the correct winding
and momentum along the circle.

In general the solution for the worldsheet fields will be
renormalized as the coupling is increased and the backreaction on the
supergravity fields is taken into account. Indeed, this is already
manifest from the general
sigma model action because, for example, $\psi_-(x^-)$
does not satisfy the fermion field equation in an arbitrary background. It is
however consistent for ${\psi}_+$ to remain zero since it appears
quadratically and quartically in the action and is thus not sourced by
any of the other worldsheet fields. Furthermore, preservation of one half of the worldsheet
supersymmetry will require that it remains zero; this follows from the supercurrents
given in (\ref{scg}). Therefore the general ansatz
for the worldsheet fields will be $(X(\s), \psi_-(\s))$, which will
be restricted further below, by fixing residual conformal symmetries.

\bigskip

The supergravity background must match the string state in symmetries,
supersymmetries and conserved quantities. In particular, it should admit
eight right moving supersymmetries and a null Killing vector, since
there are no right moving excitations. It should also describe a
solitonic string.

The most natural guess is
that the appropriate supergravity solution is a
general chiral null model \cite{Horowitz:1994rf, Tseytlin:1996yb}, namely
\bea
ds^2 &=& H^{-1}(x,v) ( - du dv + K(x,v) dv^2 + 2 A_{i} (x,v) dx^i dv) +
dx^{i} dx_{i}; \nn \\
B_{u v} &=& \half H^{-1}(x,v); \hsp B_{v i}  = H^{-1} (x,v) A_i (x,v)
;
\hsp \phi
= -\half \ln(H (x,v)). \la{bac} \eea
The supergravity equations of motion, including the worldsheet sources, are
\bea
D_{I} (e^{-2 \phi} H^{IJK}) &=& - \frac{4\k^2}{\sqrt{-g}} \frac{\d}
  {\d B_{JK}} \left (S_{\s} \d^{10}(x - X(\s)) \right )  ; \\
T^{IJ} = (R^{IJ} + 2 D^{I} D^J \phi - \qu H^{IKL}H^{J}_{\sp KL})
&=&  \frac{2 \k^2 e^{2\phi}}{\sqrt{-g}} \frac{\d}{\d g_{IJ}}
\left ( S_{\s} \d^{10} (x- X(\s)) \right ), \nn
\eea
whilst the dilaton equation of motion has no worldsheet sources
\be
4 D^2 \phi - 4 (D \phi)^2 + R - \frac{1}{12} H^2 = 0.
\ee
In the bulk of the spacetime, away from the sources, these
reduce to the following three equations
\bea \la{eeq}
T^{u v } &=& - 2 \pa^2 H = 0; \\
T^{u i} &=& 2 (\pa_{j} F^{ji} + \pa_v \pa_i H) = Y^i = 0; \nn \\
T^{u u} &=& 2 (- H \pa^2 K - K \pa^2 H + A_i Y^i + 2 H \pa_v (\pa_i A^i
- \pa_v H) ) = 0, \nn
\eea
where $F_{ij} = \pa_{i}A_{j} - \pa_{j} A_i$ and $\pa^2 = \pa_{i}
\pa^{i}$.
These equations result from those terms in the Einstein equations
which are not already automatically satisfied by the ansatz; the first
two equations also follow from the three form equations
\footnote{These equations differ from those given in
\cite{Lunin:2001jy,Lunin:2001fv} for generalized chiral null models
in which $H$ depends on $v$; these papers do not include the
terms depending on the $v$ derivatives of $H$. However, their explicit
solutions solve the equations given here rather than the equations given
there. Their gauge field $A_i$ also manifestly satisfies the gauge
condition $\pa_i A^i = \pa_v H$ rather than the Lorentz gauge condition.}.
(They coincide because of the structure $g_{vI} = B_{vI}$.) The most
familiar solitonic solutions (reviewed below)
are those for which $H$ is independent of $v$ and the
Lorentz gauge $\pa_{i} A^i = 0$ is chosen so that $H$ and $K$ are
harmonic and $F$ is coclosed. In particular, when only $H$ is
non-vanishing, the solution is such as to describe a static fundamental string.
We will however be interested in more general solutions. Note for later use that the
form of the $T^{ui}$ and $T^{uu}$ equations suggests that the natural gauge choice
when $H$ depends on $v$ is $\pa_{i} A^{i} = \pa_v H$.

Also, as pointed out in \cite{Horowitz:1994rf}, all but the zero mode of $K$ can be removed
by a coordinate transformation. That is, if $u \rightarrow (u + \chi(v,x))$ so that
$du \rightarrow (du + \pa_v \chi dv + \pa_i \chi dx^i)$, the form of the background is preserved,
but with
\be \la{ginvv}
K \rightarrow (K - \pa_v \chi), \hsp
A_i \rightarrow (A_i - \pa_i \chi).
\ee
By suitable choice
of $\chi(v,x)$ one can remove all but the zero mode of $K$. This invariance is manifest in our
explicit solutions; only the $v$ independent piece of $K$ will carry physical information.

We should also briefly comment about the exactness of these backgrounds. When there
are no source terms, the supergravity equations of motion reproduce the conditions for
exact conformal invariance of the worldsheet theory to all orders in perturbation theory
\cite{Horowitz:1994rf}. Thus, for example, gravitational wave solutions
(for which there are no sources) are exact. In this paper, however, we are interested in
including fundamental string sources and therefore the backgrounds are no longer exact.
One does have to justify the self-consistency of the supergravity approximation and we will
return to this issue at the end of the paper.

Regardless of the specific choices of the functions and
the gauge field the background preserves one quarter of the supersymmetry.
This follows from the supersymmetry variations
for the dilatino and gravitino
\bea
\d \lambda &=& (\g^{I} \pa_{I} \phi \eta^{\ast} - \frac{1}{6} H_{IJK} \g^{IJK} \eta); \\
\d \Psi_{I} &=& (\pa_{I} \eta + \qu (\w_{I}^{ab} \g_{ab} \eta - H_{I}^{ab} \g_{ab} \eta^{\ast})),
\eea
where $\eta = (\ep_1 + i \ep_2)$ is the complex Majorana-Weyl spinor
of type IIB. One can easily show that the
background (\ref{bac}) admits eight spinors satisfying $\ep_1 =
e^{\half \phi} \ep_{1}^0$  where $\ep_{1}^0$
is constant and satisfies the null projection condition
$\g^{\hat{v}} \ep_{1}^0 = 0$.

Thus the background (\ref{bac}) satisfies the two primary requirements
for it to correspond to solitonic string solutions: it has a null
isometry and the requisite supercharges. We will now discuss the
matching between fundamental string sources and the functions
appearing in the metric.

\subsection{String sources} \la{ss21}

We have reduced the coupled worldsheet and spacetime equations to
the problem of (i) finding
solutions of the worldsheet field equations and constraints and then
(ii) showing that these consistently give rise to the required source terms
in the spacetime field equations.
To solve the coupled supergravity and worldsheet equations
for the corresponding string soliton, we will proceed by making various
assumptions which will be justified by being able to self-consistently
solve the coupled equations. The first assumption we make is, as
already mentioned, that the
classical worldsheet solution has $\psi_{+}^I = 0$; that is,
these fields are not renormalized as the coupling is increased. This
is a reasonable and rather weak assumption, because it guarantees that
the classical solution preserves worldsheet supersymmetries.

The remaining worldsheet field equations (\ref{f1}) and (\ref{b1})
are consistently solved provided that
\bea
(\pa_{+} \psi_{-}^I + \G^{I(-)}_{JK} \pa_{+} X^J \psi_{-}^K) = 0; \\
(\pa_{+} \pa_{-} X^I + \G^{I(-)}_{JK} \pa_{+} X^{J} \pa_{-} X^{K}) =
0. \nn
\eea
The non-vanishing components of the torsionful connection are
collected in appendix \ref{apb}; in particular, it is important to note that
$\G^{I(-)}_{u J} = 0$. This means that the worldsheet field equations
are automatically
satisfied by any $V(x^-)$, $X^i(x^-)$ and $\psi_{-}^I(x^-)$ with $U$
the only coordinate that depends on $x^+$. Using the residual conformal
symmetry one can then fix the lightcone coordinates to be
\be \la{aq1}
U = k x^+ + {\cal U}(x^-); \hsp
V = l x^-,
\ee
where $(k,l)$ are as yet arbitrary constants and ${\cal U}$ is an undetermined
function. This choice corresponds to a static gauge in which
the solitonic string wraps the lightcone directions.
Note that the general form of this worldsheet solution is
unchanged from that in the flat background.

The sources in the supergravity equations induced by these worldsheet
fields are then
\be \la{b11}
\frac{\pi \a'} {\k^2} T^{uI} = \int d^2 \s (\pa_{+} U) \left ( 2
(\pa_{-} X^I)  \d(x^I - X^I) + i \pa_{K} (\psi_-^I \psi_{-}^K \d(x^M -
X^M)) \right ),
\ee
where the derivative is with respect to the spacetime coordinate,
rather than the worldsheet field, namely $\pa_{K} = \pa/\pa x^K$.

\section{Bosonic condensates} \la{ch3}

Let us first review earlier discussions \cite{Dabholkar:1995nc, Lunin:2001fv}
for purely bosonic worldsheet
condensates, such that $\psi_{-}^I = 0$. Because of bulk
diffeomorphism invariance, the most general bosonic solitonic string
can be generated from a worldsheet solution in which the transverse
coordinates $X^i$ are chosen to be a constant, fixed by translational
invariance to zero.
The only non vanishing components of the constraint equations are
respectively then
\be
0 = g_{vv} (\pa_{+} V)^2 + 2 g_{vu} (\pa_{+} V) (\pa_{+} U) =
g_{vv} (\pa_{-} V)^2 + 2 g_{vu} (\pa_{-}V) (\pa_{-} U),
\la{ab}
\ee
which need to be satisfied only on (and in the neighborhood of) the
worldsheet itself, that is, at $X^i = x^i = 0$. Now to solve these
equations one needs the behavior of the metric components, but these
in turn are determined by the sources, so one has to solve all
equations simultaneously and consistently. Since there are no sources
for the $T^{ui}$ equation, the background has $A_i = 0$ and $H$
independent of $v$; both $H$ and $K$ should then be harmonic away from
the sources, with the source terms being
\bea \la{x1}
\pa^2 H &=& - \frac{\k^2}{\pi \a'} \int d^2\s (\pa_+ U) (\pa_- V)
\d(x^M - X^M); \\
(H \pa^2 K + K \pa^2 H) &=& \frac{\k^2}{\pi \a'} \int d^2 \s (\pa_+ U)
(\pa_- U) \d(x^M - X^M). \nn
\eea
A generic harmonic function $T(x,v)$ on $R^{d-2}$ (with coordinates $x$)
can be decomposed into spherical harmonics $Y_l$ as
\be
T(x,v) = \sum_{l \ge 0} (a_l(v) x^l + b_l(v) x^{(2 - d - l)}) Y_l.
\ee
Terms that go as $x^{\b}$ with $\b = 0$ can be removed by coordinate
transformations; requiring that the dilaton approaches zero at
infinity (note that implicitly $g_s = 1$) fixes the constant term in
the function $H(x,v)$ to be one. The constant term in $K(x,v)$ can be removed by a
coordinate transformation and indeed must be for the metric to be
asymptotically flat; therefore we set it to zero from the start.

Terms with $\b \ge 2$ give rise to a metric which is not
asymptotically flat even after coordinate transformations. Terms with
$\b \le (1-d)$ do not contribute to conserved (monopole) charges such as the
mass and do not match onto string sources involving only bosonic
worldsheet fields. They will however play an important role in later
discussions of the fermionic condensates. The two terms which are
distinguished here are the $\b =1$ and $\b = (2-d)$ terms; the latter
are associated with the mass and momentum of the string.

Now let us take an ansatz that the harmonic functions of (\ref{bac}) behave as
\be
H = \left (1 + \frac{Q}{\left | x \right |^{(d-2)}} \right ); \hsp
K =  F(v) \cdot x, \la{sol}
\ee
where the string is localized in $d$ transverse dimensions labelled by
$x$. We have allowed the string to be only partially
localized in the transverse space for greater generality; implicitly
we are assuming $d \ge 3$ so that the resulting spacetime will be asymptotically flat.

With these choices the leading order terms in (\ref{ab}) are
$(\pa_{+} V \pa_+ U) = (\pa_- V \pa_- U) = 0$ and are satisfied
provided that
\be
U = k x^+; \hsp
V= l x^- \la{wssol}
\ee
with $(k,l)$ as yet still arbitrary.
Note that with the ansatz for the background both $g_{vv}$ and $g_{u v}$ actually
vanish on the worldsheet and so the equations (\ref{ab}) would
still be satisfied for arbitrary $U = U(x^+) + U(x^-)$, $V =
V(x^-) + V(x^+)$. However, the equations should also be satisfied within
the neighborhood of this hypersurface, as we will discuss further
below. Thus, even though $g_{uv} \sim x^2 \rightarrow 0$, if we solve
the equations in the neighborhood of $x=0$ the possible $U$ and $V$
are restricted to $U(x^+)$ and
$V(x^-)$ only which using the residual conformal invariance can be fixed to
(\ref{wssol}).

We still have to justify the ansatz for the harmonic functions, and
this follows from the source equations (\ref{x1}). In the first
equation of (\ref{x1}) is a source which matches the solution for $H$
in  (\ref{sol}). Given this behavior for $H$, the second equation is
well-defined and non-singular only if $K$ vanishes on the
worldsheet. Following the discussion above, this restricts $K$ to be
linear in $x$, since we expect that the solution should be
asymptotically flat.

The matching of the
parameters in the solutions (\ref{sol}) and (\ref{wssol}) is then
\be \la{q}
k = (\mu + R n_1); \hsp l = R n_1; \hsp
Q = \frac{n_1 \k^2}{\pi \a' (d-2) \w_{d-1}},
\ee
where the string wraps $n_1$ times around the spacetime circle of radius
$R$ as the worldsheet coordinates goes from 0 to $2 \pi$. The constant
$\mu$ is given by
\be
\mu = \frac{1}{2\pi} \int^{2 \pi R n_1}_0 dv (\pa_{v} f)^2,
\ee
where $F^i = 2 \pa_{v}^2 f^i$ and $(\pa_{v}f)^2 \equiv (\pa_v f) \cdot (\pa_v f)$.
In the form (\ref{sol}) the supergravity solution is not
asymptotically flat, but it can be brought into asymptotically flat
form by the coordinate transformations
\be
v \rightarrow v; \hsp x^i \rightarrow (x^i - f^i); \hsp
u \rightarrow u - 2 \pa_{v} f \cdot (x- f) - \int^{v} (\pa_v' f))^2
dv'.
\ee
This gives
\bea \la{rsc}
ds^2 &=& - H^{-1} du dv + (H^{-1} - 1 ) (\pa_{v} f)^2
dv^2 + 2 (H^{-1}-1) \pa_{v} f \cdot dx dv +
dy^{i} dy_{i}; \nn \\
B_{u v} &=& \half (H^{-1} - 1); \hsp B_{v i}  = \pa_{v} f_i (H^{-1}
- 1) ; \\
e^{-2 \phi} &=& H = \left (1 + \frac{Q}{\left | x - f \right |^{d-2}}
\right ), \nn
\eea
in which coordinates it is clear that the metric describes a wrapped
oscillating string. Note that this background is also a chiral null
model of the more general kind, in which $H$ depends implicitly on $v$
and the gauge field is non-zero, satisfying the gauge condition
$\pa_{i} A^{i} = \pa_{v} H$. For later use, consider the
asymptotic behavior of the $g_{vv}$
component of the metric in (\ref{rsc}), which determines the momentum
in the $v$ direction. Fourier expanding $g_{vv}$, one finds that the
zero mode term as $x \rightarrow \infty$ is
\be \la{mom1}
- \frac{\a' n_p Q}{n_1 R^2 \left | x \right |^{d-2}} = - \frac{n_p
  \k^2}{\pi R^2 (d-2) \w_{d-1} \left | x \right |^{d-2}}.
\ee
Given that the background is asymptotically flat, one can write down an energy momentum tensor
defined with respect to a flat background; this is done in \cite{Dabholkar:1995nc} and we will
not repeat the details of this discussion here.
The mass per unit length of string is then identified as $n_1/2 \pi
\a'$ whilst the lightcone momentum per unit length is $- n_p/ 2\pi
R^2$.

In the transformed coordinates the solution for the string
worldsheet fields is
\be
V = R n_1 x^-; \hsp U = (\mu + Rn_1 ) x^+ + \int^v dv' (\pa_v' f)^2; \hsp
X^i = f^i(v). \la{q1}
\ee
One can repeat the previous discussion to show explicitly how the Virasoro
constraints, the worldsheet equations and the supergravity equations
with sources are satisfied, but this of course is implied
automatically by the previous analysis.

\bigskip

These worldsheet fields also satisfy the field equations
and Virasoro constraints in the flat space limit.
Using the standard
mode expansions for fields in flat space, one can write down a
classical solution for a string wrapping a spacetime circle, carrying momentum
along this circle and carrying left moving excitations as
\bea
V = n_1 R x^-; \hsp
U = (n_1 R +  \a' \frac{n_p}{R}) x^+ + \a' \frac{n_p}{R} x^- + U'(x^-);
\hsp
X^i = f^i(x^-) \equiv f^i(v),  \la{q2}
\eea
where $n_1$ is the winding number around the circle and $n_p$ is the
integral momentum around the circle. The lightcone coordinates $v$ and
$u$ are related to the time and circle coordinates by $v = (t + y)$ and
$u = (t - y)$. Both $f^i(x^-)$ and
$U'(x^-)$ are arbitrary functions periodic in $x^-$. Imposing the
Virasoro constraints, and relating (\ref{q1}) and (\ref{q2}),
one finds that
\be
\mu = \frac{\a' n_p}{R} = \frac{1}{2 \pi} \int^{2\pi R n_1}_{0} dv
(\pa_v f)^2; \hsp
\int^{v} (\pa_{v'} f)^2 dv' = \mu x^- + U' (x^-).
\ee
Note that the original solution (\ref{wssol}) also manifestly
satisfies both the classical worldsheet field equations and the constraints in
flat space.

From the form of (\ref{q2}), one can see more specifically what quantum states the supergravity
solutions relate to; they are of the form (\ref{st}) with
\be \la{zlq}
\prod_{l} (\a^{u}_{-m_{u_l}})^{n_{u_l}}
(\a^{i_l}_{-m_{i_l}})^{n_{i_l}}
\left | e; k; n_1; n_p \right >_{NS}.
\ee
That is, they involve left moving excitations in the $u$ and $x^i$
directions, with the former fixed by
the Virasoro constraints in terms of the latter once the excitation
number is given.So roughly speaking
the non-zero coefficients in the mode expansions of the $f^i$
correspond to the oscillators $\a^{i}_{-m_{i_l}}$ appearing
in the state, with the magnitude depending on $n_{i_l}$. As already
mentioned this is not a statement that can in general be made precise,
since there is not a one to one correspondence between quantum states
and classical vibrational profiles. We will return to the matching in
\S \ref{ch7} when we discuss the regime of validity of the
supergravity approximation.

Finally, before moving on to supergravity solutions with fermionic worldsheet sources,
we should point out a subtlety suppressed in the above discussion. (Our review was aimed at
illustrating the basic ideas involved in matching worldsheet sources to supergravity fields
before moving on to rather more complicated cases. Thus we suppressed various subtleties
so as not to obscure the main points.) Our discussion
has followed that of the original paper \cite{Dabholkar:1995nc} but in a more recent paper
\cite{Lunin:2001fv} the implications of the string winding in the spacetime were
discussed. In the above discussion the solution is written in such a way that it is
not manifest that the string has $n_1$ strands: the solution is effectively written
in a covering space for $v$ in which the string has only one strand rather than in
the physical space where the supergravity fields are expressed
as a sum over these strands. This issue was the subject of
\cite{Dabholkar:1995nc}. We will not repeat their discussion here, because later we will discuss
in detail its generalization to our solutions. Suffice to say that their multi-strand solution
for the case of purely bosonic excitations can be obtained from our more general solution
of (\ref{fx11}) by setting the fermionic sources to zero.

\section{Fermionic condensates} \la{ch4}

Now we proceed to more general worldsheet field condensates,
$(X(\s),\psi_-(\s))$. Let us start with quantum states built from
left moving fermionic oscillators only and no transverse bosonic
oscillators. That is, the relevant quantum states are
\be
\prod_{l} (\a^{u}_{-m_{u_l}})^{n_{u_l}} \b^{I_l}_{r_l} \left | e;
k; n_1; n_p \right >_{NS},
\ee
where the total number of fermionic
oscillators is even (and of course each fermionic oscillator
occurs only once). It is quite straightforward to use the standard
expressions for the superconformal generators to show that there
are such physical states, with requisite momentum and winding
number, provided that one chooses the $u$ excitations
appropriately. We will demonstrate this explicitly with
corresponding classical worldsheet solutions. That is, following
the logic of (\ref{q2}) suppose that (before taking into account
backreaction on the supergravity fields) these states relate to
classical worldsheet solutions in flat space of the form
\be
\la{p11}
U = (n_1 R + \frac{\a' n_p}{R}) x^+ + \frac{\a' n_p}{R}
x^- + U'(x^-); \hsp V = n_1 R x^-; \hsp \psi_{-}^I(x^-),
\ee
with
$x^{i} = 0 = \psi_{+}^{I}$,  $U'(x^-)$ an arbitrary periodic left
moving function and $\psi_{-}^{I}(x^-)$ an arbitrary anti-periodic
left moving function. The vanishing of the right moving fermions
is manifest since the quantum state contains no corresponding
right moving fermionic oscillators. Note that the bosonic ansatz
already enforces the winding and momentum number conditions on the
$y$ circle necessary to match those of the quantum state.

The classical superconformal constraints from $T_{--}$ and $J_{-}$
then impose the conditions
\bea
\a' n_1 n_p + n_1 R \pa_{-} U' + \half i \eta_{IJ} \psi_{-}^I
\pa_{-} \psi_-^J = 0; \la{sc1} \\
( \frac{\a' n_p}{R} + \pa_{-} U') \psi_{-}^v + n_1 R   \psi_{-}^u =
0.
\eea
For later use, let us note that if we eliminate $\psi_{-}^u$ then the $T_{--}$
constraint can be rewritten as
\be \la{rewrite}
(\pa_v U) = \frac{1}{2 n_1 R} (i \psi_{-}^i \pa_v \psi_{-}^i) (1 + \frac{i}{2n_1 R}
\psi_{-}^v \pa_v \psi_{-}^v).
\ee
A generic left moving antiperiodic fermion can be expanded in modes as
\be
\psi^{I}_- = \sum_{r \in \half Z} b^{I}_r e^{i r x^-},
\ee
where the requisite reality conditions are $b^{I}_{-r} =
(b^{I}_r)^{\ast}$ and since this is now a classical solution all
$b^{I}_r$ anticommute. Fermion bilinears are necessarily periodic in $x^-$;
in particular, the combination appearing in (\ref{sc1}) is
\be
\half i \eta_{IJ} \psi_{-}^I
\pa_{-} \psi_-^J = - \eta_{IJ} \sum_{r,n} r b^{I}_{n-r} b^{J}_r
e^{i n x^-},
\ee
where $r \in \half Z$ and $n \in Z$. The zero mode constraint in
(\ref{sc1}) gives
\be \la{one}
\a' n_1 n_p =  \eta_{IJ} \sum_{r} r b^{I}_{-r} b^{J}_{r},
\ee
whilst the non-zero mode constraints give
\be \la{two}
\pa_{-} U'  = \frac{1}{n_1 R} \eta_{IJ} \sum_{r,n \neq 0} r
b^{I}_{n-r} b^{J}_{r} e^{i n x^-}.
\ee
The physical meaning of these constraints is that constant part of the
fermionic condensate matches the left moving excitation number, that is,
the product of the winding and momentum numbers. If the
fermionic bilinear condensate is not constant, then a bosonic left
moving excitation is induced by the remaining Virasoro
constraints. The first equation (\ref{one}) corresponds to the
vanishing of $L_0$ whilst the vanishing of all terms in (\ref{two})
corresponds to the classical vanishing of all $L_n$.

The second equation in (\ref{sc1}) imposes a further constraint on the components
of the fermions along the lightcone directions; this constraint relates to the unfixed
residual superconformal symmetry. That is, the natural fermion gauge choice corresponding
to the gauge choice for $V$ is to set $\psi_{-}^v = 0$; we are effectively in lightcone gauge.
Then the second equation in (\ref{sc1}) implies that $\psi_{-}^u = 0$.
More generally, suppose we consider states which also have a transverse left moving bosonic excitation
$X^{i}(x^-)$. Then the classical superconformal constraints imply that
\bea \la{scgen}
\a' n_1 n_p + n_1 R \pa_{-} U - \pa_{-} X^{i} \pa_{-}X^i + \half i \eta_{IJ} \psi_{-}^I
\pa_{-} \psi_-^J = 0; \\
( \frac{\a' n_p}{R} + \pa_{-} U) \psi_{-}^v + n_1 R  \psi_{-}^u  - \psi_{-}^i \pa_{-} X^{i} =
0.
\eea
In this case, one can still impose the gauge choice $\psi_{-}^v = 0$; the lightcone fermion $\psi_{-}^u$ is then
non-dynamical, being determined in terms of the transverse fermions and bosons, as we expect in a lightcone gauge.
Therefore, for a general (large) left moving excitation number, we find that there are eight chiral bosons and eight chiral
fermions worth of solutions to the classical constraint equations. If we freeze excitations in four transverse
directions, keeping in mind making contact with the D1-D5 system via dualities on a $T^4$, then we have four
chiral bosons and four chiral fermions worth of solutions.

\bigskip

Now let us consider the backreacted supergravity solutions with sources.
We consider first the case where the string is located at a constant
position in the transverse space, which by translational invariance is
fixed to be the origin $x^i = 0$. We defer the general discussion in which transverse bosons
are also excited to later. We have already demonstrated that
the worldsheet equations of motion in the curved background
are automatically satisfied by any $V(x^-)$, $\psi_-^I(x^-)$ with $U$
the only coordinate depending on $x^+$. Using the residual conformal
invariance we fix the lightcone coordinates as in (\ref{aq1}). The
resulting sources (\ref{b11}) are then
\bea
T^{uu} &=& \frac{\k^2}{\pi \a'} \int d^2 \s (\pa_{+} U)
\left ( 2 (\pa_-U) \d(x^M - X^M) + i (\pa_{v} (\psi^{u}_{-} \psi^{v}_{-}) +
\psi^{u}_{-} \psi^{i}_{-} \pa_{i}) \d(x^M - X^M) \right ); \nn \\
&=&  \frac{\k^2}{\pi \a' n_1 R} \left ( (2 n_1 R \pa_v {\cal U}(v) + i \pa_v
(\psi_-^u(v) \psi_{-}^v(v)) ) \d^8(x) + i \psi_-^u(v) \psi_-^i(v) \pa_{i}
(\d^8(x)) \right ); \nn \\
T^{uv} &=& \frac{\k^2}{\pi \a'} \int d^2 \s (\pa_{+}U) \left ( 2 (\pa_- V)
\d(x^M - X^M) + i \psi^{v}_- \psi^{i}_{-} \pa_{i} (\d(x^M - X^M))
\right ); \la{so1} \\
&=& \frac{\k^2} {\pi \a' n_1 R} \left ( 2 (n_1 R) \d^8(x) + i \psi^v_{-}(v)
\psi^{i}_-(v) \pa_{i} (\d^8(x)) \right ); \nn \\
T^{ui} &=&  \frac{i \k^2}{\pi \a'} \int d^2 \s (\pa_{+} U) \left (
(\pa_{v} (\psi^{i}_{-} \psi^{v}_{-}) +
\psi^{i}_{-} \psi^{k}_{-} \pa_{k}) \d(x^M - X^M) \right ) ; \nn \\
&=& \frac{i \k^2}{\pi \a' n_1 R} \left ( \pa_v
(\psi_-^i(v) \psi_{-}^v (v)) ) \d^8(x) + \psi_-^i(v) \psi_-^k(v) \pa_{k}
(\d^8(x)) \right ). \nn
\eea
Here the dependence of the sources on the lightcone coordinate is also made
explicit. These
expressions are implicitly written in terms of a lightcone coordinate $v$ which
runs between $0$ and $2 \pi n_1 R$, that is, in the covering space. In
the physical space, however, there are $n_1$ strands of a multiwound string
and the sources should be written as a sum over the strands. It is
more convenient to solve the equations in the covering space, and then
rewrite the result at the end in the physical space. (There are
various subtleties involved in this, to which we will return later.)
Note also that we have not imposed a lightcone gauge choice $\psi_{-}^v = 0$;
we have left the residual superconformal gauge invariance unfixed.

\bigskip

The main new feature of the fermionic sources is that they involve
derivatives of Dirac delta functions. Take first the $T^{uv}$ equation which
determines the function $H(x,v)$:
\be
\pa^2 H = - \frac{\k^2} {\pi \a' n_1 R } \left ( (n_1 R) \d^8(x) + \half i \psi^v_{-}(v)
\psi^{i}_-(v) \pa_{i} (\d^8(x)) \right ).
\ee
The first term on the righthandside corresponds to the $l=0$, $\b =
(2-d)$ term in the harmonic function, and relates to the mass of the
solitonic string. The second term, however, corresponds to a source
for the $l=1$, $\b = (1-d)$ term in the harmonic function, and will
give a dipole moment at infinity. Explicitly solving the equation one
finds
\be \la{hsol}
H = \left ( 1  +  \frac{\k^2}{\pi \a' (d-2) \w_{d-1}} (1  +
\frac{1}{2 n_1 R} i \psi^{v}_-
\psi^{i}_- \pa_{i}) \frac{1}{ \left | x \right |^{d-2}}  \right ).
\ee
The constant term is as usual required for the metric to be asymptotically
flat and for the string coupling to approach a constant at infinity.
The $T^{ui}$ equation reduces to
\be
(\pa_{j} F^{ji} + \pa_{v} \pa_{i} H) = \frac{i \k^2}{2 \pi \a' n_1 R} \left ( \pa_v
(\psi_-^i(v) \psi_{-}^v (v)) ) \d^8(x) + \psi_-^i(v) \psi_-^k(v) \pa_{k}
(\d^8(x)) \right ).
\ee
As mentioned after (\ref{eeq}) the most natural gauge choice is
$\pa_{i} A^{i} = \pa_{v} H$, in which case one gets
\be
\pa^2 A^{i} = \frac{i \k^2}{2 \pi \a' n_1 R} \left ( \pa_v
(\psi_-^i(v) \psi_{-}^v (v)) ) \d^8(x) + \psi_-^i(v) \psi_-^k(v) \pa_{k}
(\d^8(x)) \right ),
\ee
which is solved by
\be \la{aisol}
A^{i} = - \frac{i \k^2}{2 \pi \a' (d-2) \w_{d-1} n_1 R} ( \pa_v
(\psi_-^i(v) \psi_{-}^v (v)) + \psi_-^i(v) \psi_-^k(v) \pa_{k})
\frac{1}{ \left | x \right |^{d-2}},
\ee
where the constant term is zero to ensure asymptotic flatness. Note
that the solution is manifestly consistent with the gauge condition $\pa_{i}
A^{i} = \pa_v H$. One might wonder why $\pa_i A^i = \pa_v H$ is the natural gauge
choice, rather than the usual covariant gauge choice, the Lorentz gauge $\pa_i A^i =0$.
This is discussed briefly in appendix \ref{ape},
the issue being that the solutions are rather more
complicated in the Lorentz gauge. The remaining equation is
\bea \la{p1}
(- H \pa^2 K - K \pa^2 H + A_i Y^i) &=&  \frac{\k^2}{2 \pi \a'}
\left ( (2 \pa_v {\cal U}(v) \right . \\
&& \hsp \left .
+ i \pa_v
(\psi_-^u(v) \psi_{-}^v(v)) ) \d^8(x) + i \psi_-^u(v) \psi_-^i(v) \pa_{i}
(\d^8(x)) \right ). \nn
\eea
Let the solution for $K$ be
\be
K = (k + k^i \pa_i) \frac{1}{\left | x \right | ^{d-2}}.
\ee
Then the $T^{uu}$ equation implies
\bea
k &=& \frac{\k^2}{2 \pi \a' (d-2) n_1 R \w_{d-1}} (2 n_1 R \pa_v {\cal U} + i \pa_v
(\psi^u_- \psi^v_-)); \\
k^i &=& \frac{\k^2}{2 \pi \a' (d-2) \w_{d-1} n_1 R} (i \psi^u_-
\psi^i_-),
\eea
along with the following constraints
\bea
2 n_1 R (2 n_1 R \pa_v {\cal U} + i \pa_{v} (\psi_{-}^u \psi_{-}^v))
+ (\pa_v
(\psi_{-}^i \psi_{-}^v))^2 &=& 0; \nn \\
i \psi_{-}^v \psi_{-}^i (n_1 R \pa_v {\cal U} + \half i \pa_{v}(\psi_{-}^u
\psi_{-}^v) ) + i n_1 R \psi_{-}^u \psi_{-}^i + \psi_{-}^k \psi_{-}^i \pa_v
(\psi_{-}^{k} \psi_{-}^v) &=& 0; \la{dsit} \\
(\psi_{-}^v \psi_{-}^i) (\psi_{-}^u \psi_{-}^k)
+ (\psi_{-}^u \psi_{-}^i) (\psi_{-}^v \psi_{-}^k) -
(\psi_{-}^{j} \psi_{-}^i) (\psi_{-}^{j} \psi_{-}^k) &=& 0. \nn
\eea
These constraints follow from the cancellation of all
non-distributional terms on the left hand side of (\ref{p1}), namely
the terms proportional to $\left | x \right |^{2-d} \d^8(x)$,
$\left | x \right |^{2-d} \pa_{i} \d^{8}(x)$,
$\pa_{i} \left | x \right |^{2-d} \d^{8}(x)$ and
$\pa_{i} \left | x \right |^{2-d} \pa_{k} \d^{8}(x)$. This is
necessary for the following reason. The string sources are not
functions, but Dirac delta functions and their derivatives. There are
many smooth functions which in appropriate limits represent
delta-functions, but the resulting spacetimes depend explicitly on the
function used to represent the delta function \cite{Geroch:1987qn}. Only in the case where
the energy momentum tensor is itself a distribution are the limiting
spacetimes the same. Therefore if the distributional string source is to
uniquely determine a spacetime the lefthand side of (\ref{p1}) must
also be distributional.

Imposing classical nilpotency of the fermions, these constraints reduce to
\bea \la{dis}
n_1 R (2 n_1 R \pa_v {\cal U} + i \pa_{v} (\psi_{-}^u \psi_{-}^v)) &=&
- (\psi_{-}^i \pa_v \psi_{-}^i) (\psi_{-}^v \pa_v \psi_{-}^v); \\
(i n_1 R \psi_{-}^u + \psi_{-}^k \pa_v
\psi_{-}^{k} \psi_{-}^v ) \psi^{i}_{-} &=& 0,
\eea
with all other terms vanishing automatically.

\bigskip

We still have to show that the superconformal constraints can be
satisfied in this background with these string sources.
The superconformal constraints $T_{--}$ and $J_{-}$ respectively become
\bea
0 &=&
- (\pa_{-} V)^2 g_{vv} - 2 g_{vu} (\pa_{-}V) (\pa_{-} U) +
\half i g_{IJ} \psi_{-}^I D_{-}^{(-)} \psi_{-}^J; \\
0 &=& g_{vv} (\pa_{-}V) \psi_{-}^v + g_{vu} ( (\pa_{-} U)\psi_{-}^v
+ (\pa_{-}V) \psi_{-}^u) + g_{vi} (\pa_{-}V) \psi_{-}^i
 - \frac{i}{12} H_{IJK} \psi_{-}^I
\psi_{-}^J \psi_{-}^K, \nn
\eea
which must be satisfied on and in the neighborhood of the worldsheet.
Expanding the metric components in the vicinity of the worldsheet, and
using the nilpotency of the fermions, one finds
\bea
- 2 g_{u v} = H^{-1} &=&  \frac{\pi \a' (d-2) \w_{d-1}}{\k^2} (
\frac{i (d-2)}{2 n_1 R} \left | x \right |^{d-4}
\psi_{-}^v \psi^{i}_{-} x^{i} + \left | x \right |^{d-2} + \cdots ); \\
g_{vv} = H^{-1} K &=&
\frac{i (d-2)}{2 n_1 R \left | x \right |^2}
 (- \psi_{-}^{u}
(\psi_{-}^i x^i) +
\psi_{-}^v (\psi_-^i x^i) (\pa_v {\cal U}
+ \frac{i}{2n_1 R}  \pa_v (\psi_{-}^u \psi_{-}^v)) + \cdots ), \nn \\
g_{vi} = H^{-1} A_{i} &=& (\frac{i (d-2)}{2 n_1 R} \frac{\psi_{-}^i
  (\psi_{-}^k x^k)}{\left | x \right |^2} - \frac{(d-2)}{4 (n_1 R)^2}
\psi_{-}^v (\psi_{-}^i x^i) \frac{\pa_v (\psi_{-}^{i}
  \psi_{-}^v)}{\left | x \right |^2} + \cdots ), \nn
\eea
where the ellipses denote subleading terms.
Given these expansions the leading terms in the superconformal constraints
are of order $x^{-1}$ and impose the conditions
\be \la{sop1}
\psi_{-}^{u} \psi_{-}^v \psi_{-}^i = 0 \hsp \hsp
(i (\pa_v {\cal U}) \pa_- V \psi_{-}^v - i \psi_{-}^u \pa_{-} V - \psi_{-}^{i}
\pa_{v} \psi_{-}^i \psi_{-}^v ) = 0.
\ee
Note that the first equation holds for all $i$.
These constraints are consistent with the solvability constraints
obtained already in (\ref{dis}) provided that
\be \la{w01}
\psi_{-}^u = g(v) \psi_{-}^v,
\ee
for any arbitrary function $g(v)$ with the conditions
\bea \la{w02}
i n_1 R g(v) + \psi_{-}^k \pa_{v} \psi_{-}^k = 0; \\
2 n_1 R \pa_{v} {\cal U} = - \frac{1}{n_1 R} (\psi_{-}^{i} \pa_v \psi_{-}^i) (\psi_{-}^v
\pa_{v} \psi_{-}^v) = i g(v)(\psi_{-}^v \pa_{v} \psi_{-}^v) .
\eea
Just as for the classical equations in flat space, the superconformal
constraints mean that $\psi_{-}^u$ and ${\cal U}$ are completely
determined by the fermions $(\psi_{-}^v,\psi_{-}^i)$. For any choices
of the latter one can solve these equations. However, for the solution to have
a given momentum $n_p$ per unit length, and hence to correspond to the
states of interest, one needs to impose a further constraint on the
zero mode of $\pa_v {\cal U}$ which reduces the freedom in the
Grassmann functions.

Furthermore, not all choices of the fermions will
produce physically distinct solutions: as already mentioned there is a residual
superconformal invariance which has not yet been fixed. This is the
reason there are nine free Grassmann functions, whilst one might have expected
only eight. Now let us discuss the reason why we have not fixed the gauge to be $\psi_{-}^v = 0$
ab initio. When $\psi_{-}^v =0$ all fermion bilinear source terms vanish except for that
in $A_i$ proportional to $\psi_{-}^i \psi_{-}^k$. In particular, this implies that the leading order term
in the solution for $A_i$ vanishes, and this in turn forces $K$ to vanish when one imposes the distributional
constraints of (\ref{dsit}). Thus the only non-vanishing terms in the solution are
the first and second terms in $H$ (\ref{hsol}) and the second term in $A_i$ (\ref{aisol}).

Thus it might seem that in this limit one obtains a
family of supergravity solutions with vanishing momentum $n_p$, non-vanishing winding number $n_1$
and non-vanishing gauge fields $A_i$. This contradicts the expectation from the flat
space spectrum, for which $n_p = 0$ automatically implies no left moving
excitations. However, there is no contradiction:
in the limit that $\psi_{-}^v = 0$ the leading terms in the
superconformal constraints (\ref{sop1}) vanish automatically
and the first non-vanishing term imposes the constraint
\be
i \psi_{-}^{i} \pa_{v} \psi_{-}^i = 0.
\ee
The zero mode of this constraint implies that
\be
\sum_{r > 1/2} r \left | b^i_{r} \right |^2 = 0,
\ee
which can manifestly only be satisfied for vanishing $b^{i}_{r}$ and hence
$\psi_{-}^i$. Thus the gauge fields have to vanish, and the
solution reduces to that for the static solitonic string with winding number $n_1$, as expected.

This still leaves a puzzle: in the flat space limit we could consistently impose the gauge $\psi_{-}^v = 0$
and obtain a family of classical solutions with non-zero $\psi_{-}^i$, $n_1$ and $n_p$. Here we only get
non-trivial solutions with winding and momentum
if both $\psi_{-}^v$ and at least one of the $\psi_{-}^i$ are non-zero. Note that this conclusion
is unaffected by choosing a different gauge choice for $A_i$. For example, in the Lorentz gauge one can
consistently solve the supergravity equations of motion, but the superconformal constraint $J_-$ cannot be satisfied.
There is one possible loophole, which is to relax the distributional constraint, and we analyse this in
appendix \ref{apad}. However, it seems unlikely that this is correct since the supergravity equations are not really
consistently solved in this case.

So within our ansatz it seems we cannot consistently impose the gauge $\psi_{-}^v = 0$. Presumably the resolution to this
puzzle is that one would need a different ansatz for the supergravity background in order to impose
$\psi_{-}^v = 0$. Recall that we are solving a complicated coupled system of supergravity and worldsheet
equations. We have found a specific
family of self-consistent solutions, which has picked out a non-zero gauge choice for $\psi_{-}^v = 0$. This does
not exclude there being a family of solutions in which one can choose $\psi_{-}^v = 0$, although from the
considerations above it seems they cannot be generalized chiral null models. From a physical perspective there
is no problem in choosing a non-zero gauge for $\psi_{-}^v$: we have enough solutions to account for all corresponding
quantum states involving fermionic excitations. The main subtlety is in the matching of the supergravity solutions
to these quantum states, and we will return to this later when discussing explicit solutions.

Before leaving this section let us note that the function ${\cal U}$ defined by the equations in
(\ref{w02}) is not an
ordinary function: nilpotency of the fermions imposes that it is also
nilpotent. There are hence implicit nilpotency restrictions on the
harmonic functions and gauge field of the supergravity background; in
particular, $K^2 = 0$. Whilst this nilpotency is unusual, it is to be expected
when the fields are sourced by fermions and does not lead to any
inconsistencies. In any case, the solutions written down so far are
not really supergravity solutions: we have to take into account the
winding, and for a valid supergravity solution we will require $n_1
\gg 1$. In this limit, as we shall see below,
the functions are only nilpotent at much higher order.

\section{Solitons in the physical space} \la{ch5}

Thus far we have worked in the covering space in which the $v$
coordinate runs between $0$ and $2 \pi n_1 R$. In the physical space
the $v$ coordinate runs between $0$ and $2 \pi R$, and the solitonic
string has $n_1$ strands. In this space the harmonic functions and
gauge field should be written as a sum over the strands
\bea \la{sx1}
H &=& \left ( 1  +  \sum_{a=1}^{n_1} \frac{\k^2}{\pi \a' (d-2) \w_{d-1}} (1  +
\frac{1}{2 n_1 R} i \psi^{a v}_-
\psi^{a i}_- \pa_{i}) \frac{1}{ \left | x \right |^{d-2}}  \right
). \\
A^{i} &=& - \sum_{a=1}^{n_1} \frac{i \k^2}{2 \pi \a' (d-2) \w_{d-1} n_1 R} ( \pa_v
(\psi_-^{ai} \psi_{-}^{av}) + \psi_-^{ai} \psi_-^{ak} \pa_{k})
\frac{1}{ \left | x \right |^{d-2}}, \nn \\
K &=&  \sum_{a=1}^{n_1}
\frac{\k^2}{2 \pi \a' (d-2) n_1 R \w_{d-1}} (2 n_1 R \pa_v {\cal U}^a
+ i \psi^{au}_- \psi^{ai}_- \pa_{i})
\frac{1}{ \left | x \right |^{d-2}}. \nn
\eea
Here the functions on the strands $F^a(v)$ are defined for $0 \le v
\le 2 \pi R$ from the function in the covering space $F(v)$ (for which
$0 \le v \le 2 \pi n_1 R$) by the relation
\be
F^a(v) = F(v + 2 \pi (a-1) R).
\ee
This implies by continuity that $F^a(0) = F^{a-1}( 2 \pi R)$ or,
stated more physically, the ends of adjacent strands are coincident
since the string is continuous. Using (\ref{mom1}), one must impose
the constraint that
\be
(\pa_{v} {\cal U})_0 = - \frac{n_{p} \a'}{n_1 R^2},
\ee
where we use the notation $(d(v))_0$ to denote the zero mode of a function
$d(v)$. This enforces the correct momentum charge of the solution.

The physically interesting case is when the number of strands (winding
charge) is large; it is only in this limit that our leading order
supergravity solution can be valid, since its curvature can be small on
the string scale. In this limit, the nilpotency of the functions
ceases to be an issue, since they only vanish at much higher order,
namely $K^{2n_1}$. In the multistrand case, the $n$-th power of a
function involves the following fermion bilinear structures:
\be
\left (\sum_{a=1}^{n_1} \psi^{aI}_- (v) \psi^{aJ}_{-} (v) \right )^n
\ee
Now $\psi^{aI}_{-} (v) \psi^{a'I}_{-} (v) $ is non-zero for generic
$a \neq a'$; there are effectively $n_1$ distinct fermions
$\psi^{aI}_{-}$ for each $I$. Expressed in terms of the original
fermion functions, $\psi^{aI}_{-} (v) \psi^{a'I}_{-} (v) $ is non-zero since
this results from multiplying $\psi^{I}_{-}( v + 2 \pi (a-1) R)
\psi_{-}^{I} (v + 2 \pi (a'-1) R)$. The latter is non-zero since the
fermionic sources are not at the same location on the worldsheet.
Thus typically the functions in the
metric are only nilpotent at the $2n_1$-th power.

\bigskip

As mentioned previously, there is an important subtlety in going from the
covering space to the physical space in
which the string is multiply wound. In the covering space,
there is a one to one correspondence between
each point on the string and each point in the target space.
In the physical space, the string is multiply wound
around the $v$ direction, so that there are $n_1$ points $x^-_a$
with $a = 1,..,n_1$ which correspond to the
same $v$. When the condensate on the string is purely fermionic,
the strands of the string are not separated
in the transverse space, and are all coincident. This does not
affect the solution of the supergravity equations of
motion around the sources. It does, however, make the solution of
the superconformal constraints rather subtle.
The general structure of the superconformal constraints is
\be \la{m12}
\sum_k G_k [x^I] \Phi_k [X^I(\s),\psi^I_-(\s)] \d(x^I - X^I) = 0,
\ee
where $G_k$ is a functional of the spacetime coordinates
(determined by the supergravity fields) and
$\Phi_k$ is a functional of the worldsheet fields and we sum over all
strands, labelled by $k$. The Dirac delta function restricts the supergravity
fields to the worldsheet. In the neighborhood of the string
worldsheet, however, the supergravity fields are dominated by the
terms
\be  \la{n13}
G_{k} = \sum_{a=1}^{n_1} \frac{G_{ka} \cdot x}{ \left | x \right |^2},
\ee
where the summation over $a$ arises from the fact that each spacetime
point $v$ corresponds to $n_1$ distinct points on the
worldsheet and thus supergravity fields get contributions from all
$n_1$ strands.
Recall that the $a$th term in this expansion is determined
by the worldsheet sources at $x^-_a$. Now if one substitutes
(\ref{n13}) into (\ref{m12}) it manifestly imposes non-local
relationships between worldsheet fields, that is, relationships
between worldsheet fields at $x^{-}_a$ and $x^-_{a'}$. Implicitly
in writing the solution (\ref{sx1}) we have solved the superconformal
constraints (\ref{m12}) only for the local terms, ie imposing
relationships between fields at the same points on the worldsheet. We
have not imposed the vanishing of non-local cross-terms
between worldsheet fields at $x^{-}_a$ and $x^-_{a'}$. This is
physically reasonable: one takes into account only the terms in the
supergravity fields sourced by that strand of the string. In any case,
the problem would be overconstrained and generically unsolvable if we
tried to impose the non-local constraints in addition.

A more rigorous way to justify our solution is the following. Suppose
we separate the individual strands in the transverse directions by
turning on a small transverse bosonic excitation in addition to the fermionic
excitation. Since the strands are now separated in the target space,
the superconformal constraints will necessarily impose local
relationships. More explicitly, at a given point $x^-_a$ on the worldsheet,
the restriction of the supergravity fields to the worldsheet will be
dominated by only one of the $n_1$ terms in the summation, that which
is sourced by the worldsheet fields at $x^-_a$. The solution of the coupled
equations for generic left moving bosonic and fermionic excitations is
discussed in detail in appendix \ref{apc} and makes this point
manifest. Since the solution with purely fermionic excitations is a
limiting case, which should be a smooth limit of the mixed condensate
case, we would expect our solution of the fermionic equations to be valid.

\section{Mixed condensates} \la{ch6}

To account for all quantum states at a given left moving excitation number we
need to switch on both fermionic and bosonic left moving excitations. As mentioned
around equation (\ref{scgen}) we expect the generic solution to be determined in
terms of eight chiral bosons and eight chiral fermions.
The BPS supergravity background corresponding to generic left moving
excitations
\be
U = (n_1 R + \frac{\a' n_p}{R}) x^+ + {\cal U}(x^-); \hsp
V = n_1 R x^-; \hsp
X^{I} = f^i(x^-); \hsp
\psi_{-}^I(x^-),
\ee
is determined in appendix \ref{apc}. The resulting multistrand solution in
the physical space is given by the following harmonic functions and
gauge field
\bea
H &=& \left ( 1 + \frac{\k^2}{2 \pi \a' \w_{d-1} (d-2)} \sum_{a=1}^{n_1}
(h_a + \td{h}^i_a \pa_i) \frac{1}{\left | x - f_a \right |^{d-2}} \right
); \nn \\
K &=& \frac{\k^2}{2 \pi \a' \w_{d-1} (d-2)} \sum_{a=1}^{n_1}
\left ( k_a + \td{k}^i_a \pa_i \right )
\frac{1}{\left | x - f_a \right |^{d-2}}; \la{fx11} \\
A_i &=& \frac{\k^2}{2 \pi \a' \w_{d-1} (d-2)} \sum_{a=1}^{n_1}
\left ( a_a^i + \td{a}^{ij}_a \pa_j \right )
\frac{1}{\left | x - f_a \right |^{d-2}}, \nn
\eea
where
\bea
h_a &=& 2; \hsp
\td{h}^i_a = \frac{i}{n_1 R} \psi_{-}^{av} \psi_{-}^{ai}; \\
k_a &=& (2 \pa_v {\cal U}^a + \frac{i}{n_1 R} \pa_v (\psi_{-}^{au}
\psi_{-}^{av}) ) ; \hsp
\td{k}^i_a = \frac{i}{n_1 R} (\psi_{-}^{au} \psi_{-}^{ai} -
\psi_{-}^{au} \psi_{-}^{av} \pa_{v} f^i_a); \nn \\
a^{i}_a &=& (-2 \pa_v f^i_a - \frac{i}{n_1 R} \pa_v (\psi_{-}^{ai}
\psi_{-}^{av})); \hsp
\td{a}^{ij}_a = \frac{i}{n_1 R} \psi_{-}^{ai} (\psi_{-}^{av} \pa_v f^j_a
- \psi_{-}^{aj}). \nn
\eea
The constraints between the v-dependent functions involved
in this solution are derived in the
appendix; again $\psi_{-}^u$ and $\cal U$ are determined by the other functions:
\bea
\psi_{-}^{au} = 2 \psi_{-}^{ai} \pa_{v} f^{i}_a + g^a \psi_{-}^{av}; \nn \\
i n_1 R g^a + (\psi_{-}^{ai} \pa_v f^{i}_a) \pa_v \psi_{-}^{av} + \psi_{-}^{ia}
\pa_{v} \psi_{-}^{ia} = 0; \\
\pa_{v} {\cal U}^a + \frac{i}{n_1 R} \psi_{-}^{ak} \psi_{-}^{av} \pa_v^2 f^k_a
= - \frac{1}{(2n_1 R)^2} (\pa_v (\psi_{-}^{ai} \psi_{-}^{av}))^2 + (\pa_v f^i_a)^2. \nn
\eea
Note that this solution manifestly reduces to the purely
bosonic and fermionic cases already given on setting $\psi_{-}^{aI}=0$ and $f^{i}_a
=0$ respectively.

The generic form for these supergravity solutions is rather complicated, since
one needs to sum over a large number of strands.
If however one considers solutions for which the mean
wavelengths are large compared to the scale of the compactification
circle, the neighboring strands give similar contributions to the
harmonic functions \cite{Lunin:2001fv, Lunin:2001jy}
and one can reasonably replace the summation over
strands with an integral.
\be
\sum_{a=1}^{n_1} \rightarrow \frac{1}{2  \pi R} \int^{2\pi R n_1}_{0} dv,
\ee
This means that the summations in (\ref{fx11}) reduce to integrals, for example
of the type
\be \la{approx}
\sum_{a} \frac{d_a(v)}{ \left | x - f_a \right |^{d-2}} \rightarrow
\frac{1}{2 \pi R} \int^{2 \pi R n_1}_0 dv \frac{d(v)}
{ \left | x - f \right |^{d-2}}.
\ee
Such an integration considerably simplifies the explicit computation
of the metric functions. In the case of purely fermionic excitations,
it is interesting to note that this approximation (of picking out the zero mode)
actually becomes exact.
The point is that in this case the sources are located at the origin in the transverse
space, and the summation over strands reduces to terms of the form
\be
\sum_a \frac{d_a(v)}{\left | x \right |^{d-2}}, \hsp
\sum_a d_a^i(v) \pa_i \frac{1}{\left | x \right |^{d-2}},
\ee
where the functions $d_a$ and $d^i_a$ are defined in terms of fermion bilinears.
By construction, each $d_a$ therefore descends from a function which is periodic on the worldsheet
and which can be expanded in integral $x^-$ harmonics. Thus,
\be
d_a(v) = \sum_{m \in Z} d_m e^{im (x^- + \frac{2 \pi a}{n_1})},
\ee
where the spacetime lightcone coordinate $v = n_1 R x^-$. Now summing over all the strands
and using the identity
\be
\sum_{a=1}^{n_1} e^{\frac{2\pi i m a}{n_1}} = 0,
\ee
valid for all $m \neq 0$, we find that
\be \la{exacferm}
\sum_{a=1}^{n_1} \frac{d_a(v)}{\left | x \right |^{d-2}}
= n_1 \frac{d_0}{\left | x \right |^{d-2}},
\ee
where $d_0$ is the zero mode. This shows that (\ref{approx}) is exact
in the purely fermionic source case; the supergravity solution depends only on the zero
modes of the worldsheet sources. Since the leading term in $A_i$ in (\ref{sx1}) has no zero modes,
this term necessarily vanishes.

\section{Matching with microstates and regularity} \la{ch7}

In the analysis so far we have solved the leading order supergravity equations of motion
with worldsheet sources, without addressing the validity of this
approximation. In this section we will discuss when such an approximation
is self-consistent. We will also discuss in more detail the matching of the supergravity
solutions with the quantum microstates, an issue touched upon in earlier sections.

Let us consider first the case of purely bosonic condensates. These have been extensively
discussed in recent literature in the context of constructing microstates for the D1-D5 system.
Here we will review issues relevant to the discussion for fermionic condensates, and furthermore
highlight certain other points which are not usually mentioned.
The prescription of \cite{Lunin:2001fv, Lunin:2001jy} for matching between quantum states and vibration
profiles is that given a string microstate
\be \la{stz1}
\prod_{l} (\a_{-n_l}^{i})^{m_l} \left | e ; k \right >_{NS},
\ee
one reads off a corresponding set of classical vibration profiles
as
\be \la{stz2}
f^{i} \sim \sum_{n_l} m_l a^i_{-n_{l}} e^{i n_l x^-},
\ee
where now $a^{i}_{-n_l}$ is a classical (commuting) quantity of order
$\sqrt{\a'}$ and appropriate reality constraints must also be
imposed. We have already pointed out that (\ref{stz1}) does not in fact give
a physical state; instead it must generically be of the form (\ref{zlq}) with
appropriate lightcone excitations included. The significance of this is that it illustrates
how the superconformal gauge choice is related to the fixing of
bulk diffeomorphism invariance.

As commented earlier it is not possible even in principle to
make a one to one correspondence between quantum states and classical
vibration profiles well defined: neither are observables and
Ehrenfest's theorem depends on using observables. Thus one must generically read
the correspondence in terms of, for example, correlation functions
computable in both limits. There are exceptions to this. For example, there is a unique
quantum state of maximal angular momentum in a given transverse plane which can be uniquely
matched to a geometry with corresponding rotational symmetry and angular momentum.
\footnote{Note that closely related issues were also considered in \cite{Iengo:2003ct} where the
semiclassical decay of certain very massive string states in flat
space was discussed.} Indeed these geometries are often used as the explicit examples in the discussions
by Mathur et al.

In making the correspondence between (\ref{stz1}) and (\ref{stz2}) one is implicitly
relating an infinite family of worldsheet conserved charges to an infinite family of
multipole moments of the spacetime geometry. It is not clear that one can make such a
correspondence well defined. The most precise way of making a correspondence between
geometries and the microscopic description is to go the dual D1-D5 frame and use the
standard AdS/CFT dictionary. That is, as mentioned in the introduction, one would read
from the asymptotics of the decoupled AdS region of the geometry the operator deformations
and vevs in the dual CFT. Then the geometries derived from purely bosonic condensates
in the F1-P system should
account for $c = 4 N = 4 n_1 n_5$ worth of R sector vacua in the CFT. Of course one cannot get
a discrete number in the continuum supergravity description, but quantizing along the lines of
\cite{Palmer:2004gu,Grant:2005qc} (counting supertube configurations and quantizing supergravity
geometries respectively) could reproduce such a number. Note however that neither approach will be
entirely self-consistent since most of the geometries one is counting actually have string
scale curvatures; we return to this issue shortly.

Now let us move on to the issue of when the supergravity approximation is valid. The curvatures are
both conformal frame and duality frame dependent, and one is usually most interested in the
corresponding geometries in the D1-D5 system. These can be obtained
from the fundamental string geometries by a series of dualities, the
explicit map being \cite{Lunin:2001jy}
\bea
&& P NS1 (IIB) \stackrel{S}{\rightarrow} P D1 (IIB)
\stackrel{T_{T^4}}{\rightarrow} P D5 (IIB)
\stackrel{S} {\rightarrow} P NS5 (IIB) \\
&& \hsp \hsp \stackrel{T_{6}} {\rightarrow} P NS5 (IIA)
\stackrel{T_5} {\rightarrow} NS1 NS5 (IIB)
\stackrel{S}{\rightarrow} D1 D5 (IIB), \nn
\eea
where $S$ and $T_{i}$ denote S and T dualities respectively, with the
latter along the $i$th direction. Here the ten-dimensional spacetime
is compactified on a four torus, with $x^6$ one of the circles in
this torus. $x^5 \equiv y$ is the spatial circle which the original string
wraps. Since one of the T-dualities is along the $y$ circle, one can only
explicitly map supergravity solutions which are $y$ independent to the
D1-D5 system; this requires that the starting geometries are independent of the $v$
direction. A generic F1-P solution depends explicitly on the
lightcone coordinate $v$, and hence $y$. However, as we have already mentioned,
the summation over many strands can be approximated as an integral over $v$ (\ref{approx}),
giving a geometry independent of $v$ which can be dualized; this approximation becomes
exact in the purely fermionic case (\ref{exacferm}).

The details of the dualisation procedure are given in \cite{Lunin:2001jy};
the resulting D1-D5 background is the (string frame) metric
\be \la{d1d5}
ds^2 = \sqrt{\frac{1}{H(1+K)}} ( -(dt -A)^2 + (dy+B)^2 )
+ \sqrt{H(1+K)} dx_i dx^i + \sqrt{\frac{(1+K)}{H}} dz_a dz_a,
\ee
with the other fields being
\be
e^{2 \phi} = \frac{(1+K)}{H}; \hsp
C_{ti} = \frac{B_i}{(1+K)}; \hsp
C_{ty} = - \frac{K}{(1+K)}; \hsp
C_{iy} = - \frac{A_i}{(1+K)}; \hsp C_{ij},
\ee
where the forms $B_i$ and $C_{ij}$ are not independent, but rather
are defined by the duals
\be
dC = - \ast d H; \hsp
dB = - \ast d A,
\ee
where the duals are taken with respect to the flat metric on
$R^4$. The two functions and gauge field follow from those in the
original F1-P system; there is a subtlety of appropriately rescaling lengths,
but the details are not important in what follows.
Note that this dualisation is equally valid for our mixed
condensate geometries in the F1-P system, since they were also expressed in
terms of generalized chiral null models.

There has been considerable discussion of the geometries in the D1-D5 system;
the main results which are relevant here follow from
\cite{Mathur:2005zp,Maldacena:2000dr,Lunin:2002iz,Lunin:2002bj,Lunin:2001jy,
Lunin:2001fv,Palmer:2004gu}. The geometries are non-singular for any
generic choice of vibration profiles $f^{i}$; one can understand this in terms of the branes
blowing up into a supertube in their transverse directions. However, the geometry will only be weakly
curved if the characteristic size of the profile is sufficiently large and the vibration profiles
are sufficiently smooth. The latter can be phrased in terms of multipole moments of the charge
distributions: high multipole moments should be small or vanishing. Generically if the vibration
profile involves small contributions from many harmonics the profile will be very fuzzy and
will not be well described by a supergravity solution. If one uses the known microscopic
distributions of states to determine the most probable vibration profiles, one finds that the generic
geometry actually has large curvatures, even though it is non-singular. One can trace this property
to the fact that the density of states is peaked around states involving many different harmonics
each of order $\sqrt{n_1 n_5}$ for which the vibration profiles are fuzzy.

For many explicit discussions, such as comparing scattering calculations with those in the CFT, certain
non-generic geometries are usually considered: these are the ones
with definite $R$ charge (angular momentum) which do not have multipole moments and for large
enough angular momentum are weakly curved everywhere. Such geometries derive from vibration profiles
which are circles in a transverse plane. In particular, the unique state with maximal angular momentum
corresponds to putting all of the excitation energy into the lowest harmonic, so that the vibration profile
is a circle of maximal radius.

Actually one should not be surprised that the generic geometry is not weakly curved. This is consistent
with the picture in which the black hole geometry emerges as a suitable averaging over these horizon
free geometries. In the supergravity approximation the black hole does not have a finite horizon, although
it is anticipated that it develops one when one evaluates $\a'$ corrections on the geometry. The characteristic
scale of this horizon must necessarily be small, in order to match with the CFT entropy. Now if this horizon
emerges as an averaging over the non-singular geometries, the latter must generically have a characteristic
scale which is small and they should not be well described by supergravity solutions.

\bigskip

We now discuss related issues for the mixed condensate geometries. We have already seen that
the matching between quantum states and geometries is rather more subtle in this case. The issue was that
the natural gauge choice $\psi_-^v = 0$ was not possible in the curved background. Presumably it would be possible
with a different ansatz for the background (related to that used here by some complicated diffeomorphism),
but to settle this one needs a deeper understanding of the relationship between worldsheet superconformal invariance
and spacetime diffeomorphism invariance.

In any case, if the dependence of the solution on $\psi_-^v$ reflects an unfixed gauge choice, then all non-zero
choices of $\psi_{-}^v$ which preserve the superconformal constraint (physically, the momentum charge) must
be physically equivalent. Demonstrating this explicitly is beyond the scope of this paper, but let us note that
shifting $\psi_{-}^v$ whilst preserving the momentum charge does not affect any other monopole charges. This follows
from the form of the supergravity solution. The shift will generically affect dipole moments, but it is not clear
whether these are physically distinguishable or simply reflect diffeomorphism invariance. Note that, as mentioned
above, one may be able to relate multipole moments of the supergravity solution to the infinite family of
conserved worldsheet charges. Thus this issue again goes back
to understanding the relationship between worldsheet symmetries and spacetime diffeomorphisms.
Again the clearest way to match geometries to a microscopic description will be in the D1-D5 system using the AdS/CFT
dictionary.

Let us move on to the question of when the supergravity approximation is self-consistent. There is a fundamental
difference between the bosonic and fermionic condensates. One can excite an arbitrary number
of quanta of any bosonic oscillator, so one can, for example, put all of the excitation energy into the lowest harmonics.
However one can only excite one quantum of each fermionic oscillator. This translates into a bound in the magnitude
of the harmonic coefficients in the classical fermion fields: each $b^{I}_r$ must be of order $\sqrt{\a'}$. To achieve
large winding and momentum we will thus need to excite a large number of harmonics; this issue is exemplified
in appendix \ref{apd}. Moreover a classical treatment of the fermion bilinear condensates can only be justified when
the scale of the bilinears is large compared to $\a'$. If it is not we cannot approximate by classical expectation
values; note also that we have treated the fermions as nilpotent, neglecting terms of order $\a'$ in their
anticommutators.

Within these constraints one may ask whether there are any simple explicit solutions, analogous to the circular
vibration profiles in the bosonic case. Suitably developing the toy example discussed in \ref{apd},
the following solution may provide such an example:
\bea
H &=& \left ( 1 + \frac{n_1 \k^2}{\pi \a' (d-2) \w_{d-1} \left | x \right |^{d-2}} \right ); \nn \\
A &=& \frac{i \k^2}{2 \pi R \w_{d-1} \a'}
(\psi_-^1 \psi_{-}^2)_{0} \frac{\cos^2 \theta}{\left | x \right |^{d-2}} d \phi; \\
K &=& - \frac{n_p \k^2}{\pi R^2 (d-2) \w_{d-1} \left | x \right |^{d-2}}
- \frac{\k^2}{2 \pi \a' \w_{d-1} R} (i \psi^{u}_- \psi^{1}_-)_0 (\frac{\cos\theta (\cos \phi - \sin \phi)}
{\left | x \right |^{d-3}}. \nn
\eea
In this expression, $d_0$ again denotes the zero mode of the function $d(v)$. We have turned on $\psi_{-}^v$,
$\psi_{-}^1$ and $\psi_-^2$; we relate the Cartesian coordinates $(x^1,x^2,\cdots)$ to polar coordinates via
$x^1 = x \cos \theta \sin \phi$, $x^2 = x \cos \theta \cos \phi$ etc. The solution has been simplified by
choosing the functional form of $\psi_{-}^1$ to be the same as that of $\psi_{-}^2$; this implies the simplified
form for the gauge field and for the dipole term in $K$. We also choose the zero mode of $\psi_{-}^v \psi_{-}^1$
to vanish, which removes the dipole term in $H$. This can be achieved by choosing the phase of $\psi_{-}^v$ to
differ from that of $\psi_{-}^1$; that is, let $\psi_{-}^v$ be an expansion in cosines, whilst $\psi_{-}^1$ is
an expansion in sines. We have already imposed the momentum charge constraint; this implies that
\be \la{modec}
2 n_p n_1 R^2 \a' = \left ( (\psi_{-}^{i} \pa_{-} \psi_{-}^i) (\psi_{-}^v \pa_{-} \psi_{-}^v) \right )_0.
\ee
Finally recall that $\psi_{-}^u$ is given by $\psi^{u}_{-} = i \psi_{-}^v (\psi_{-}^i \pa_- \psi_{-}^i)/(n_1 R)^2$.

As an explicit example of how the constraints can be satisfied
choose solutions for the fermions which involve exciting one unit of all harmonics up to
some given level. That is,
\be
\psi_{-}^v = \sqrt{\a'} \sum_{r = 1/2}^{r^v/2} \ep^v_r \cos (r x^-); \hsp
\psi_{-}^{i} = \sqrt{\a'} \sum_{r = 1/2}^{r^{i}/2} \ep^{i}_{r} \sin (r x^-),
\ee
where the dimensionless coefficients $\ep^{I}_r$ anticommute and have magnitude of order one. Now assume that
$r^{i} \gg r^v \gg 1$. The momentum charge constraint then enforces
\be
n_p n_1 (\frac{R^2}{\a'}) \sim (r^i)^2 (r^v)^3,
\ee
which can be satisfied by choosing $r^i \sim \sqrt{n_p n_1}$ and $(r^v)^3 \sim R^2/\a'$. This
is manifestly self-consistent in the limit $\sqrt{n_p n_1} \gg (R/\sqrt{a'})^{1/3} \gg 1$ which
is within the interesting parameter range. Note that the form of this constraint can be worked out by
approximating the convolution of the harmonics in the quartic fermion term of (\ref{modec}) using the
form of the fermion fields. By construction
our solution also has non-zero angular momentum in the $\phi$ direction; the scale of this angular momentum
(per unit length of string) is set by
\be
J_{\phi} \sim \frac{(\psi^1_- \psi^2_-)_{0}}{R \a'} \sim \sqrt{n_p n_1}/R,
\ee
where the latter estimate is obtained by approximating the strength of the zero mode harmonic in
the convolution of the quadratic fermion term. The solution also has dipole charges arising from the
subleading harmonics. So, to summarize, we have constructed an explicit F1-P solution with winding $n_1$
and momentum $n_p$ and non-zero angular momentum in one transverse plane, along with subleading dipole
moments. It is interesting to note that if we substitute the same fermion solution into the flat space
superconformal constraints (\ref{rewrite}) they are also consistently satisfied by the same choices of
$r^i$ and $r^v$. Moreover, the conserved worldsheet angular momentum charge $J^{12}$ is
\be
J^{12} \sim \frac{1}{\a'} \int d\s (\psi^1_- \psi^2_-) \sim \sqrt{n_p n_1},
\ee
which matches the conserved charge of the curved spacetime. All other angular momentum charges of the
worldsheet theory vanish, also in agreement with those in the spacetime. Thus there is at least
approximate matching between the spacetime and the quantum state (obtained naively from the classical
solution, lifting classical harmonic coefficients to the excitations of the quantum state).

\bigskip

Now let us consider the dual of this solution in the D1-D5 frame. Restricting
to $d=4$ and then putting the explicit forms for the functions and gauge fields
into (\ref{d1d5}) one finds the following metric in the decoupled AdS region:
\bea
ds^2 &=& \frac{x^2}{\l(x^i)} ( -(dt -A)^2 + (dy + B)^2 )
+ \l(x^i) (\frac{dx^2}{x^2} + d \Omega_3^2 + dz_a \cdot dz_a); \\
\l(x^i) &=& \sqrt{1 + \a x^{-1} \cos \theta (\cos \phi - \sin \phi)},
\hsp
A_{\phi} = \frac{\b \cos^2 \theta}{x^2},
\hsp
B_{\chi} = \frac{\b \sin^2 \theta}{x^2}, \nn
\eea
where $d \Omega_3^2 = (d \theta^2 + \cos^2 \theta d\phi^2 + \sin^2 \theta d\chi^2)$
is the metric on the unit three sphere. For simplicity we have suppressed most scale
factors, including the AdS scale derived from the D1 and D5 brane numbers,
retaining only the novel terms $\a \sim (\psi_{-}^u \psi_{-}^1)_0$ and
$\b \sim (\psi^1_1 \psi^2_-)_0$. There are two reasons we focus on the near horizon (decoupled)
region. The main one is that the singularities of the spacetime are clearly confined to this
region, and become manifest in the decoupled limit. The second is that this would provide
the starting point for matching the AdS asymptotics to the dual CFT.
This solution illustrates what is probably a generic
property of geometries corresponding to purely fermionic condensates: it
is singular as $x \rightarrow 0$, the intuitive reason being that the branes have not blown up
in the transverse directions. This specific geometry also has closed timelike curves confined to
the region $x < \b$ and is singular as $\l(x^i) \rightarrow 0$ which occurs for $x > 0$.
It is an open question whether the generic fermion condensate and mixed condensate
geometries are non-singular, as the bosonic condensate geometries are.

\section{Conclusions and Discussion} \la{conc}

The motivation for this paper was the observation that there are insufficient two
charge geometries to account for all microscopic states in the F1-P system and its
duals. The known geometries are
characterized by four chiral bosons, which upon quantization can account for
$c = 4 n_1 n_p$. (We know from the analysis in the F1-P system that each chiral
boson gives $c = n_1 n_p$ when one enforces the momentum and winding charge constraints.)
But one would expect the more general geometry to be characterized by four
chiral bosons along with four chiral fermions, which upon quantization can give
the full $c = 6 n_1 n_p$ required to account for all microstates. In this paper we have
constructed the missing geometries and explored some of their properties.

Throughout the paper we have raised a number of issues which merit further investigation.
In particular, one needs to understand the relationship between worldsheet superconformal
invariance and spacetime diffeomorphism invariance; a related issue is whether
one can unambiguously relate
the infinite number of conserved charges of the worldsheet theory to spacetime multipole
moments. One needs to show when and whether the geometries sourced by fermionic condensates
can be well described within the supergravity approximation. One should also explore the matching
between the geometries and their dual CFT descriptions using the AdS/CFT dictionary.

Many other interesting issues have not so far been raised in this paper. For example, it is conjectured
that the bosonic condensate geometries can be understood in terms of supertubes. Presumably our solutions
may also admit a description in terms of supertubes carrying fermionic condensates, and such a
description may prove useful. An obvious question is the generalization to three charge geometries
for which the corresponding black hole has a macroscopic horizon. Unlike the two charge system, the
generic bosonic condensate geometry is not known; only certain specific families of geometries have been
constructed \cite{Lunin:2004uu,Bena:2005va}.
Perhaps one can also find at least some explicit geometries corresponding to operators
in the dual CFT built from fermions, extending the ideas in this paper. The relation to the bubbling
picture \cite{Lin:2004nb}, extended to the D1-D5 system in
\cite{Boni:2005sf,Bena:2005va}, would also be interesting to explore -
at least some of the mixed condensate geometries
should preserve enough R symmetry to be included by the bubbling description. Note however that
the bubbling description is best understood for geometries which are regular in supergravity, whereas
these geometries do not seem to be regular.

Even if turns out that all two charge fermion condensate solutions require a description which goes
beyond supergravity, one cannot ignore them in the context of the Mathur conjecture. They are needed
to account for the full microscopic entropy; indeed this should be demonstrated by looking at the
AdS asymptotics, and showing that geometries of the type given here are required to account for
the CFT ground states built from fermionic operators.

A few years ago it would have been considered a hopeless problem to understand the stringy resolution
of such geometries, despite their eight supercharges. This generically requires a knowledge
of all $\a'$ corrections to the string effective action. However, recent developments suggest that the
stringy resolution might be more under control than one would have expected. For example, several families
of BPS black holes seem to admit non-renormalization theorems, in that only a subset of all possible
corrections are needed in order to account for their microscopic entropy \cite{BH}.
So far this is well understood only in one case, that where the black hole horizon contains an $AdS_3$ factor
\cite{Kraus:2005vz}.
If the Mathur conjecture is correct, then a black hole of this type emerges from an averaging over our horizon free
geometries, so perhaps the corrections on these geometries are also heavily constrained.
Another more speculative possibility is that one can extract from the bubbling description geometries beyond
the supergravity approximation; this would entail understanding in more detail how the free fermion configuration
determines the (exact) dual geometry.

\section*{Acknowledgments}

This work is supported by NWO under the Vidi grant
``Holography, duality and time dependence in string theory''.

\appendix

\section{Sigma model action} \la{apa}

The sigma model action before gauge fixing is \cite{Bergshoeff:1985qr}
\bea
S_{\s} &=& \frac{1}{2 \pi \a'} \int d^2 \s \left ( - \half \sqrt{-\g}
\g^{\mu \nu} \pa_{\mu} X^I \pa_{\nu} X^J g_{IJ}
- \half \ep^{\mu \nu} B_{IJ} \pa_{\mu} X^{I} \pa_{\nu} X^J \right . \nn \\
&& \hsp - \half \sqrt{-\g} i \bar{\psi}^{I} \g^{\mu} (\pa_{\mu}
\psi^{J} + \G^{J}_{\sp KL} \pa_{\mu} X^{K} \psi^{L}) g_{IJ} \\
&& \hsp + \sqrt{-\g} (\bar{\chi}_{\mu} \g^{\nu} \g^{\mu} \psi^{I}
\pa_{\nu} X^{J} g_{IJ} - \qu \bar{\chi}_{\mu} \g^{\nu}
\g^{\mu} \chi_{\nu} \bar{\psi}^{I} \psi^{J} g_{IJ} - \frac{1}{12}
R_{IJKL} \bar{\psi}^{I} \psi^{K} \bar{\psi}^J \psi^L) \nn \\
&& \hsp + \sqrt{-\g} ( - \qu i H_{IJK} \bar{\psi}^I \g^{\mu} \g_{5} \psi^{J} \pa_{\mu} X^{K}
+ \frac{1}{12} i H_{IJK} \bar{\chi}_{\mu} \g^{\nu} \g^{\mu} \psi^{I}
\bar{\psi}^{J}
\g_{\nu} \g_{5} \psi^{K}) \nn \\
&& \hsp + \sqrt{-\g} \left . (\frac{1}{16} D_{K} H_{ILJ}
\bar{\psi}^{I} \psi^{K}
\bar{\psi}^{J} \g_{5} \psi^{L} - \frac{1}{32}
g^{MN} H_{IKM} H_{JLN} \bar{\psi}^{L} \g_{5} \psi^{K} \bar{\psi}^J \g_{5} \psi^L) \right ), \nn
\eea
where $\g_{\mu \nu}$ is the worldsheet metric and $\chi_{\mu}$ is the
worldsheet gravitino. The worldsheet spinors are two
component Majorana. In coordinates $(x^0,x^1)$, the flat 2d metric is
taken to be $\eta_{\mu \nu} = \rm{diag} (-1,1)$ and the 2d gamma matrices are
$\g^{0} =\s_2$, $\g^1 = i \s_1$ and $\g_{5} = - \s_3$. We take
$\ep^{01} = - \ep_{01} = -1$ and $\bar{\chi} = \chi^{t} \g^0$.
In lightcone coordinates $x^{\pm} = (x^0 \pm x^1)$, the flat metric
becomes $\g_{+-} = - \half$ and $\ep^{+-} = 2$ with $\ep_{+-} = -
\half$. The positive and negative chirality components of the spinors
are defined by $\psi_{\mp} = \half (1 \mp \g_5) \psi$. Upon gauge fixing the worldsheet
metric to be flat and the gravitino to vanish, the action reduces to
that given in the main text.

\section{Curvature of chiral null background} \la{apb}

It is convenient to define the torsionful connections for this
background as
\be
\G^{I (\pm)}_{\sp JK} \equiv (\G^{I}_{\sp JK} \pm \half H^{I}_{\sp
  JK}),
\ee
such that
\be
\G^{I (-)}_{\sp JK} = \G^{I (+)}_{\sp KJ} = \half g^{IL} (\pa_{J}
C_{KL} + \pa_{K} C_{IL} - \pa_{L} C_{IJ}),
\ee
where $C_{IJ} = g_{IJ} + B_{IJ}$. The non-vanishing components are
then given by
\bea
\G^{v(-)}_{\sp vi} &=& - \pa_{i} (\ln H); \hsp \G^{i(-)}_{\sp vu} =
\half \pa^{i} H^{-1}; \hsp
\G^{i(-)}_{\sp vv} = H^{-1} \pa_{v} A^{i} - \half \pa^{i} (H^{-1} K);
\\
\G^{i (-)}_{v j} &=& - A_{j} \pa^{i} H^{-1} - H^{-1} F^{i}_{\sp j}; \hsp
\G^{u (-)}_{i u} = - \pa_{i} (\ln H); \hsp
\G^{u (-)}_{v u } = A^{i} \pa_{i} H^{-1}; \nn \\
\G^{u(-)}_{v v } &=& - \pa_{v} K + 2 H^{-1} A^{i} \pa_{v} A_{i} - A^{i}
\pa_{i} (H^{-1} K) - H^{-1} K (\pa_v H); \hsp
\G^{v (-)}_{vv } = - H^{-1} \pa_v H; \nn \\
\G^{u (-)}_{iv} &=& - \pa_{i} K + K \pa_{i} (\ln H); \hsp
\G^{u (-)}_{ij} = -2 \pa_{i} A_j + 4 A_{j} \pa_{i} (\half \ln(H)); \nn
\\
\G^{u (-)}_{vi} &=& - \pa_{i} K - K \pa_{i} (\ln H) - 2 A_{i} A^{j}
\pa_{j} (H^{-1}) + 2 H^{-1} A^{j} F_{ij}. \nn
\eea
Note in particular that $\G^{I (-)}_{uJ} = 0$.

\section{Solutions with $\psi_{-}^v = 0$ } \la{apad}

In this appendix we explore further whether solutions with $\psi_{-}^v = 0$ can be found
self-consistently within our ansatz. First we impose $\psi_{-}^v$ in the source terms
given in (\ref{so1}). We then solve the $T^{uv}$ and $T^{ui}$ equations as before, to give
\bea
H &=& \left ( 1  +  \frac{\k^2}{\pi \a' (d-2) \w_{d-1} \left | x \right |^{d-2}} \right ); \\
A^{i} &=& - \frac{i \k^2}{2 \pi \a' (d-2) \w_{d-1} n_1 R} (\psi_-^i(v) \psi_-^k(v) \pa_{k})
\frac{1}{ \left | x \right |^{d-2}}. \nn
\eea
Note that $\pa_v H = 0$ and $A^i$ satisfies the Lorentz gauge condition. The final supergravity
equation is
\be \la{vgy}
(- H \pa^2 K - K \pa^2 H)  = \frac{\k^2}{2 \pi \a'} (2 \pa_v {\cal U}(v)) \d^8(x).
\ee
All other terms on the left vanish, due to the Lorentz gauge and the vanishing of $A_{i} \pa_{f} F^{ji}$.
Matching the delta function terms gives the previous solution
\be
K = \frac{\k^2 \pa_v {\cal U} }{\pi \a' (d-2) \w_{d-1} \left | x \right |^{d-2}}.
\ee
However, there is a non-distributional term on the left of (\ref{vgy}) which does not cancel; thus
in the main text we imposed the vanishing of $\pa_v {\cal U}$. Suppose we relax this constraint,
and simply proceed to solve the superconformal constraints. These can now be consistently solved
for non-vanishing $\psi_{-}^i$ since there are two contributing terms in the $T_{--}$ constraint.
\be
0 = (- g_{vv} (\pa_{-}V)^2 + \half i \psi_{-}^{i} \pa_{-} \psi_{-}^{i},
\ee
(implicitly this is evaluated in the neighborhood of the string at $x \rightarrow 0$)
which is solved by setting
\be
(n_1 R)^2 \pa_v {\cal U} = \half i \psi_{-}^i \pa_{-} \psi_-^i.
\ee
This clearly reduces to the same equation given in the main text if one imposes the vanishing of
$\pa_{v} {\cal U}$, but if this constraint is relaxed one can find a family of geometries with
fixed winding and momentum, characterized by eight independent Grassmann functions with
$\psi_{-}^v = 0$. Moreover the worldsheet solutions and superconformal constraint manifestly
match those for the string in flat space. However, it is not clear that dropping the distributional
constraint is justified; after all, the supergravity equation (\ref{vgy}) is not consistently solved.
This would be true no matter what function we choose to represent the delta function. Moreover, the
solutions would differ from the naive N1-P solution only when at least two of the fermions are non-zero.

This issue
clearly requires more in depth analysis, but for the purposes of this paper let us note the following
points. The solutions we give in a non-zero gauge for $\psi_{-}^v$ are clearly self-consistent and
do (non-trivially) reduce to previously known solutions for purely bosonic excitations when one
sets the fermion terms to zero. Furthermore, the generic
properties of our supergravity backgrounds are the same as those of the above background,
so many of the discussions about curvature singularities, conserved charges etc would also be applicable to
this case.

\section{Solitons carrying bosonic and fermionic condensates} \la{apc}

In this appendix we consider the supergravity equations and
superconformal constraints for solitonic strings carrying bosonic and
fermionic condensates. In this case the worldsheet fields are
\be
U = n_1 R x^+ + {\cal U}(x^-); \hsp
V = n_1 R x^-; \hsp
X^{i} = f^{i}(x^-); \hsp
\psi^{I}_{-} (x^-),
\ee
which automatically satisfy the worldsheet equations of motion in any
chiral null background since $\G^{I(-)}_{u K} = 0$. The sources for
the supergravity equations of motion are given by (\ref{so1}) (with
the obvious replacement $\d^{(8)}(x) \rightarrow \d^{(8)}( x - f)$)
with the additional terms
\bea
\d T^{uu} &=& \frac{\k^2}{n_1 R \pi \a'} (i \psi_{-}^u \psi_{-}^v)
(-\pa_v f^i) \pa_i \d^{(8)}(x^k - f^k) ; \\
\d T^{ui} &=& \frac{\k^2}{\pi \a'} \left (2 (\pa_v f^i) \d^{(8)}(x^k - f^k)
- \frac{i}{n_1 R} \psi_{-}^i \psi_{-}^v (\pa_{v} f^{k}) \pa_{k}
(\d^{(8)}(x^j - f^j)) \right ). \nn
\eea
These terms give rise to the following additional terms in the
harmonic function $K$ and the gauge field:
\bea
\d K &=& \frac{ \k^2}{2 \pi \a' (n_1 R) (d-2) \w_{d-1}} (i
  \psi_{-}^u \psi_{-}^v \pa_{v}) \frac{1}{\left | x - f \right
    |^{d-2}}; \\
\d A^{i} &=& \frac{\k^2}{2 \pi \a' (d-2) \w_{d-1}} \left ( -
\frac{2 \pa_v f^i}{\left | x - F \right |^{d-2}} + i \frac{\psi_{-}^i
  \psi_{-}^v}{n_1 R} (\pa_v f^k) \pa_{k} \frac{1}{\left | x - f \right
  |^{d-2}} \right ). \nn
\eea
Note that the original terms in the harmonic functions and gauge field also get
shifted by $x \rightarrow (x-f)$.

We now need to write down the constraints imposed by the supergravity
energy-momentum tensor being distributional and the superconformal
constraints. Starting with the $J_-$ constraint, this implies that
\be
(\psi_{-}^{u} - 2 \psi_{-}^i \pa_v f^i) \psi_{-}^v \psi_{-} \cdot  (x
- f) = 0,
\ee
which is solved provided that
\be
(\psi_{-}^{u} - 2 \psi_{-}^i \pa_v f^i) = g(v) \psi_{-}^v,
\ee
for some function $g(v)$. Notice that solving this equation first
eliminates $\psi_{-}^u$ and simplifies the other equations.

The superconformal constraint for $T_{--}$ implies that
\be \la{hy1}
(i n_1 R (g(v) - \pa_v U) + \psi_{-}^k \pa_v f ^k \pa_v
\psi_{-}^v  + 2 i n_1 R
(\pa_v f)^2 + \psi_{-}^i \pa_v \psi_{-}^i) \psi_{-}^v = 0.
\ee
The constraints arising from the vanishing of terms in the energy
momentum tensor proportional to
\be
\pa_{i} \frac{1}{\left | x - f \right |^{d-2}} \d^{(8)}(x^k - f^k);
\hsp
\frac{1}{\left | x - f \right |^{d-2}} \pa_{i} \d^{(8)}(x^k - f^k),
\ee
are identical and imply that
\be \la{hy2}
0 = (i n_1 R (\pa_v {\cal U} + g(v)) + \psi_{-}^k \pa_v f^k \pa_v
\psi_{-}^v + \psi_{-}^k \pa_v \psi_{-}^k).
\ee
The remaining constraint from the energy momentum tensor comes
from the vanishing of terms proportional to $\left | x - f \right |^{2-d} \d^{(8)}(x-f)$
and gives
\be \la{hy3}
\pa_{v} {\cal U} + \frac{i}{n_1 R} \psi^{k}_{-} \psi_{-}^v \pa_v^2 f^k
= - \frac{1}{(2 n_1 R)^2} (\pa_v (\psi_{-}^i \psi_{-}^v))^2 + (\pa_v
f)^2.
\ee
These three constraints (\ref{hy1}), (\ref{hy2}) and (\ref{hy3}) are
consistently solved provided that
\be
(i n_1 R g(v) + (\psi_- \cdot \pa_v f) \pa_v \psi_{-}^v + \psi_{-}^i
\pa_v \psi_{-}^i) = 0,
\ee
with (\ref{hy3}) also satisfied. To check the consistency of this
solution with (\ref{hy1}) and (\ref{hy2}) first
note that (\ref{hy3}) implies via the nilpotency of the fermions that
\be
(\pa_{v} {\cal U} - (\pa_v f)^2) \psi_{-}^v = 0.
\ee
Then multiply (\ref{hy2}) on the right with $\psi_{-}^v$ and subtract
(\ref{hy1}); this gives precisely the equation above.

\section{Gauge choices} \la{ape}

In this appendix we discuss why the gauge choice $\pa_i A^i = \pa_v H$ is natural.
In all the cases discussed here the relevant terms in the $T^{ui}$ and $T^{uu}$ equations are
of the form
\bea \la{apz1}
2 (\pa_j F^{ji} + \pa_v \pa_i H) = Y^i = y^{i}(v) \d^{8}(x - f) + \cdots ; \\
2 (-H \pa^2 K - K \pa^2 H + A_i Y^i + 2 H \pa_v (\pa_i A^i - \pa_v H)) = 2 k \d^{8}(x -f) + \cdots,
\eea
with the $T^{uv}$ equation determining the function $H$.
The ellipses denote the additional source terms in the fermionic case, which do not
play a role in this discussion. We consistently solved these equations with the gauge
choice $\pa_{i} A^i = \pa_v H$ for both fermionic and mixed condensates. That this is a natural
gauge choice is manifest from the first of these equations:
\be
2 (\pa_{j} F^{ji} + \pa_v \pa_i H) \rightarrow 2 \pa^2 A^i = y^{i}(v) \d^{8}(x-f) + \cdots,
\ee
and thus each component of the gauge field is harmonic. The consistency of this solution with the
gauge choice is guaranteed in all of our examples by the relationship between the sources for
$H$ in the $T^{uv}$ equation and those in the $T^{ui}$ equation.

Suppose we instead imposed the usual Lorentz gauge choice $\pa_{i} \hat{A}^{i} = 0$. The solutions to the
equations (\ref{apz1}) are then rather more complicated, the first equation being
\be
2 (\pa^2 \hat{A}^i + \pa_v \pa_i H) = y^i(v) \d^8(x-f) + \cdots,
\ee
which is solved by $\hat{A}^i = A^i + a^i$ with
\be
a^i = - (\pa^2)^{-1} \pa_v \pa_i H.
\ee
In particular, the gauge field components are no longer harmonic functions. Similarly, from the second equation
in (\ref{apz1}) one immediately sees that $K$ is no longer harmonic since it also picks up extra terms.
Indeed the shift in $K$ is implicit from equation (\ref{ginvv}), defining $a^i = - \pa_{i} \chi$.
Thus the gauge used throughout the paper is the most natural and simplest.

\section{An explicit fermionic example} \la{apd}

To demonstrate how the equations and constraints of \S \ref{ch4} can be solved, let us
consider a particular toy example. Switch on the lowest harmonics of
the three worldsheet fermions $(\psi_{-}^v, \psi_{-}^{1}, \psi_{-}^2)$
so that
\bea
\psi_{-}^v &=& \left (b^v_{1/2} e^{i \half x^-} + b^{v}_{- 1/2} e^{- i
  \half x^-} \right ); \nn \\
\psi_{-}^1 &=& \left (b^1_{1/2} e^{i \half x^-} + b^{1}_{- 1/2} e^{- i
  \half x^-} \right );  \\
\psi_{-}^2 &=& \left (b^2_{1/2} e^{i \half x^-} + b^{2}_{- 1/2} e^{- i
  \half x^-} \right ), \nn
\eea
where recall that the reality conditions imply that $b^{I}_{r} =
(b^{I}_{-r})^{\ast}$. Solving the constraints of (\ref{w01}) and
(\ref{w02}) gives
\bea
\psi_{-}^{u} = g(v) \psi_{-}^v = \frac{1}{(n_1 R)^2} \left (\left
| b^1_{1/2} \right |^2 + \left
| b^2_{1/2} \right |^2 \right ) \psi_{-}^v; \\
\pa_{v} {\cal U} = \frac{1}{2 (n_1 R)^4} (\left
| b^1_{1/2} \right |^2 + \left
| b^2_{1/2} \right |^2)  \left | b^v_{1/2} \right |^2.
\eea
Then note that each of the fermion bilinears can be written as
\be
i \psi_{-}^{aI} \psi_{-}^{aJ} = 2 {\rm{Re}} (i b_{1/2}^{I}
(b^J_{1/2})^{\ast}) + 2 {\rm{Re}} (i b^{I}_{1/2} b^{J}_{1/2}
e^{ix^-_a}),
\ee
with $x^{-}_a = v/(n_1 R) + 2 \pi (a-1)/n_1$. Using the identity
\be
\sum_{a=1}^{n_1} e^{2 \pi (a-1) i/n_1} = 0,
\ee
one finds that
\be
i \sum_{a=1}^{n_1} \psi_{-}^{aI} \psi_{-}^{aJ} = 2 n_1 {\rm{Re}} (i b_{1/2}^{I}
(b^J_{1/2})^{\ast}).
\ee
Thus the explicit forms for the harmonic functions and gauge field are
\bea
H &=& \left (1 + \frac{n_1 \k^2}{\pi \a' (d-2) \w_{d-1}}
\left (1 + \frac{1}{n_1 R} ({\rm{Re}} (i b^{v}_{1/2} (b^1_{1/2})^{\ast})
\pa_1  + {\rm{Re}} (i b^{v}_{1/2} (b^2_{1/2})^{\ast})
\pa_2) \right ) \frac{1}{\left | x \right |^{d-2}} \right ); \nn \\
A_i &=& - \frac{\k^2}{\pi \a' (d-2) \w_{d-1} R}
\left ( {\rm{Re}} (i b^{1}_{1/2} (b^2_{1/2})^{\ast})
\pa_2  + {\rm{Re}} (i b^{2}_{1/2} (b^1_{1/2})^{\ast})
\pa_1) \right ) \frac{1}{\left | x \right |^{d-2}}; \\
K &=& \frac{\k^2}{\pi \a' (d-2) \w_{d-1} (n_1 R)^2 R}
\left ( \frac{1}{n_1 R} (\left
| b^1_{1/2} \right |^2 + \left
| b^2_{1/2} \right |^2)  \left | b^v_{1/2} \right |^2 \right .
\nn \\
&& \hsp \hsp \hsp \left . + \left | b^2_{1/2} \right |^2
{\rm{Re}} (i b^{v}_{1/2} (b^1_{1/2})^{\ast})
\pa_1
+ \left | b^1_{1/2} \right |^2 {\rm{Re}} (i b^{v}_{1/2} (b^2_{1/2})^{\ast})
\pa_2 \right  ) \frac{1}{\left | x \right |^{d-2}}. \nn
\eea
This example has a number of interesting features. The supergravity
background is independent of $v$, since the $v$ dependence
cancelled when we summed over the strands, as shown in (\ref{exacferm}).
Moreover, the leading order term in the gauge field
cancelled; this guarantees that the metric is asymptotically flat in
the usual sense. The leading order term in $K$ is nilpotent, again
since we switched on only the lowest harmonic in $\psi_{-}^v$. With
mixed harmonics in $\psi_{-}^v$ $(\pa_{v} {\cal U})$ does not have to
nilpotent, as we discussed earlier. Note that in this example we need
to switch on two transverse fermions for the gauge field to be
non-vanishing; if we set $\psi_{-}^2$ to zero the gauge field
vanishes.

One can see quite easily that this specific example does not really
give a good supergravity solution: it is only a toy example.
Following from (\ref{mom1}) one can
read off the momentum of the solitonic string from the asymptotics of
the function $K$; this gives
\be \la{i1}
n_p = - \frac{ (\left | b^1_{1/2} \right |^2 + \left
| b^2_{1/2} \right |^2)  \left | b^v_{1/2} \right |^2 }{\a' n_1^3 R^2}.
\ee
Given the normalizations in the worldsheet action,
the worldsheet fermions have dimension one and thus the $b^{I}$
are dimensionful, in units of $(\a')^{1/2}$.
Now in the expansion in harmonics
of a bosonic excitation, the (dimension one) coefficients of each harmonic
\be
X^{I}_{n} = \sum_{n \ge 0} a^{I}_{n} \cos(n x^-) + \sum_{n > 0}
\bar{a}^{I}_{n} \sin (n x^-),
\ee
can be arbitrarily large. The size of the coefficient essentially relates
to the number of quanta of that harmonic in the quantum state, which
is of course unbounded. Now consider the fermionic excitation expanded
in harmonics; here the coefficients $b^{I}_{r'}$ cannot be arbitrarily
large, since one can only excite one fermionic quantum of each
harmonic. Therefore, $b^{I}_{1/2}$ is necessarily of order
$(\a')^{1/2}$ and thus
\be
n_p \sim \frac{1}{n_1^3 (R/\sqrt{\a'})^2}.
\ee
This is clearly outside the validity of the supergravity approximation, in
which one requires $n_p \gg 1$, $n_1 \gg 1$ and the radius $R$ to
be large compared to the string scale. The issue is that one cannot
achieve large winding and momentum charges by exciting only a few
quanta of the lowest harmonics of the fermions! This is in contrast to
the bosonic case, where one can of course put all of the excitation energy into
the lowest harmonics.


\begin{thebibliography}{99}

\bibitem{Gauntlett:2002nw}
  J.~P.~Gauntlett, J.~B.~Gutowski, C.~M.~Hull, S.~Pakis and H.~S.~Reall,
  ``All supersymmetric solutions of minimal supergravity in five dimensions,''
  Class.\ Quant.\ Grav.\  {\bf 20}, 4587 (2003)
  [arXiv:hep-th/0209114].

\bibitem{Dabholkar:1989jt}
  A.~Dabholkar and J.~A.~Harvey,
  ``Nonrenormalization Of The Superstring Tension,''
  Phys.\ Rev.\ Lett.\  {\bf 63} (1989) 478.

\bibitem{Dabholkar:1990yf}
  A.~Dabholkar, G.~W.~Gibbons, J.~A.~Harvey and F.~Ruiz Ruiz,
  ``Superstrings And Solitons,''
  Nucl.\ Phys.\ B {\bf 340}, 33 (1990).

\bibitem{Dabholkar:1995nc}
  A.~Dabholkar, J.~P.~Gauntlett, J.~A.~Harvey and D.~Waldram,
  ``Strings as Solitons \& Black Holes as Strings,''
  Nucl.\ Phys.\ B {\bf 474}, 85 (1996)
  [arXiv:hep-th/9511053].

\bibitem{Sen}
A.~Sen,
``Extremal black holes and elementary string states,''
Mod.\ Phys.\ Lett.\ A {\bf 10}, 2081 (1995)
[arXiv:hep-th/9504147].

\bibitem{Callan:1995hn}
  C.~G.~Callan, J.~M.~Maldacena and A.~W.~Peet,
  ``Extremal Black Holes As Fundamental Strings,''
  Nucl.\ Phys.\ B {\bf 475}, 645 (1996)
  [arXiv:hep-th/9510134].

\bibitem{Maldacena:1997re}
  J.~M.~Maldacena,
  ``The large N limit of superconformal field theories and supergravity,''
  Adv.\ Theor.\ Math.\ Phys.\  {\bf 2}, 231 (1998)
  [Int.\ J.\ Theor.\ Phys.\  {\bf 38}, 1113 (1999)]
  [arXiv:hep-th/9711200].

\bibitem{Gubser:1998bc}
  S.~S.~Gubser, I.~R.~Klebanov and A.~M.~Polyakov,
  ``Gauge theory correlators from non-critical string theory,''
  Phys.\ Lett.\ B {\bf 428}, 105 (1998)
  [arXiv:hep-th/9802109].

\bibitem{Witten:1998qj}
  E.~Witten,
  ``Anti-de Sitter space and holography,''
  Adv.\ Theor.\ Math.\ Phys.\  {\bf 2}, 253 (1998)
  [arXiv:hep-th/9802150].

\bibitem{Mathur:2005zp}
  S.~D.~Mathur,
  ``The fuzzball proposal for black holes: An elementary review,''
  arXiv:hep-th/0502050.


\bibitem{Lunin:2001jy}
  O.~Lunin and S.~D.~Mathur,
  ``AdS/CFT duality and the black hole information paradox,''
  Nucl.\ Phys.\ B {\bf 623}, 342 (2002)
  [arXiv:hep-th/0109154].

\bibitem{Lunin:2001fv}
  O.~Lunin and S.~D.~Mathur,
  ``Metric of the multiply wound rotating string,''
  Nucl.\ Phys.\ B {\bf 610}, 49 (2001)
  [arXiv:hep-th/0105136].

\bibitem{Lunin:2002bj}
  O.~Lunin, S.~D.~Mathur and A.~Saxena,
  ``What is the gravity dual of a chiral primary?,''
  Nucl.\ Phys.\ B {\bf 655} (2003) 185
  [arXiv:hep-th/0211292].

\bibitem{Maldacena:2000dr}
  J.~M.~Maldacena and L.~Maoz,
  ``De-singularization by rotation,''
  JHEP {\bf 0212} (2002) 055
  [arXiv:hep-th/0012025].

\bibitem{Lunin:2002iz}
  O.~Lunin, J.~Maldacena and L.~Maoz,
  ``Gravity solutions for the D1-D5 system with angular momentum,''
  arXiv:hep-th/0212210.

\bibitem{deBoer:1998ip}
  J.~de Boer,
  ``Six-dimensional supergravity on S**3 x AdS(3) and 2d conformal field
  theory,''
  Nucl.\ Phys.\ B {\bf 548}, 139 (1999)
  [arXiv:hep-th/9806104].

\bibitem{Maldacena:2000hw}
  J.~M.~Maldacena and H.~Ooguri,
  ``Strings in AdS(3) and SL(2,R) WZW model. I,''
  J.\ Math.\ Phys.\  {\bf 42}, 2929 (2001)
  [arXiv:hep-th/0001053].


\bibitem{Bergshoeff:1985qr}
  E.~Bergshoeff, S.~Randjbar-Daemi, A.~Salam, H.~Sarmadi and E.~Sezgin,
  ``Locally Supersymmetric Sigma Model With Wess-Zumino Term In Two-Dimensions
  And Critical Dimensions For Strings,''
  Nucl.\ Phys.\ B {\bf 269}, 77 (1986)

\bibitem{Das:1989wf}
  A.~K.~Das, J.~Maharana and S.~Roy,
  ``The Neveu-Schwarz-Ramond String In Background Fields: Nilpotency Of Brst
  Charge,''
  Nucl.\ Phys.\ B {\bf 331}, 573 (1990).

\bibitem{Das:1988zk}
  A.~K.~Das, J.~Maharana and S.~Roy,
  ``Brst Quantization Of Superstring In Backgrounds,''
  Phys.\ Rev.\ D {\bf 40}, 4037 (1989).


\bibitem{Horowitz:1994rf}
  G.~T.~Horowitz and A.~A.~Tseytlin,
  ``A New class of exact solutions in string theory,''
  Phys.\ Rev.\ D {\bf 51}, 2896 (1995)
  [arXiv:hep-th/9409021].

\bibitem{Tseytlin:1996yb}
  A.~A.~Tseytlin,
  ``Generalised chiral null models and rotating string backgrounds,''
  Phys.\ Lett.\ B {\bf 381}, 73 (1996)
  [arXiv:hep-th/9603099].

\bibitem{Geroch:1987qn}
  R.~Geroch and J.~H.~Traschen,
  ``Strings And Other Distributional Sources In General Relativity,''
  Phys.\ Rev.\ D {\bf 36}, 1017 (1987).


\bibitem{Iengo:2003ct}
  R.~Iengo and J.~G.~Russo,
  ``Semiclassical decay of strings with maximum angular momentum,''
  JHEP {\bf 0303}, 030 (2003)
  [arXiv:hep-th/0301109]; D.~Chialva, R.~Iengo and J.~G.~Russo,
  ``Decay of long-lived massive closed superstring states: Exact results,''
  JHEP {\bf 0312}, 014 (2003)
  [arXiv:hep-th/0310283].

\bibitem{Palmer:2004gu}
  B.~C.~Palmer and D.~Marolf,
  ``Counting supertubes,''
  JHEP {\bf 0406} (2004) 028
  [arXiv:hep-th/0403025].

\bibitem{Grant:2005qc}
  L.~Grant, L.~Maoz, J.~Marsano, K.~Papadodimas and V.~S.~Rychkov,
  ``Minisuperspace quantization of 'bubbling AdS' and free fermion droplets,''
  arXiv:hep-th/0505079.

\bibitem{Lunin:2004uu}I.~Bena and P.~Kraus, ``Three charge supertubes and black hole hair,'' Phys.\ Rev.\ D {\bf 70} (2004) 046003
  [arXiv:hep-th/0402144];
  O.~Lunin, ``Adding momentum to D1-D5 system,'' JHEP {\bf 0404} (2004) 054
  [arXiv:hep-th/0404006]; S.~Giusto, S.~D.~Mathur and A.~Saxena, "Dual geometries for a set of 3-charge microstates,"
  Nucl.\ Phys.\ B {\bf 701} (2004) 357 [arXiv:hep-th/0405017]; S.~Giusto, S.~D.~Mathur and A.~Saxena,
  ``3-charge geometries and their CFT duals,'' Nucl.\ Phys.\ B {\bf 710} (2005) 425 [arXiv:hep-th/0406103];
 I.~Bena and P.~Kraus, ``Microstates of the D1-D5-KK system,'' Phys.\ Rev.\ D {\bf 72} (2005) 025007
  [arXiv:hep-th/0503053].

\bibitem{Bena:2005va}
  I.~Bena and N.~P.~Warner,
  ``Bubbling supertubes and foaming black holes,''
  arXiv:hep-th/0505166.


\bibitem{BH} K.~Behrndt, G.~Lopes Cardoso, B.~de Wit, D.~Lust, T.~Mohaupt and W.~A.~Sabra,
Phys.\ Lett.\ B {\bf 429}, 289 (1998) [arXiv:hep-th/9801081];
G.~Lopes Cardoso, B.~de Wit and T.~Mohaupt,
Phys.\ Lett.\ B {\bf 451}, 309 (1999)
[arXiv:hep-th/9812082]; G.~Lopes Cardoso, B.~de Wit and T.~Mohaupt,
Fortsch.\ Phys.\  {\bf 48}, 49 (2000)
[arXiv:hep-th/9904005]; G.~Lopes Cardoso, B.~de Wit and T.~Mohaupt,
Nucl.\ Phys.\ B {\bf 567}, 87 (2000)
[arXiv:hep-th/9906094]; G.~Lopes Cardoso, B.~de Wit and T.~Mohaupt,
Class.\ Quant.\ Grav.\  {\bf 17}, 1007 (2000)
[arXiv:hep-th/9910179];
G.~Lopes Cardoso, B.~de Wit, J.~Kappeli and T.~Mohaupt,
JHEP {\bf 0012}, 019 (2000)
[arXiv:hep-th/0009234]; A.~Dabholkar,
arXiv:hep-th/0409148; A.~Dabholkar, R.~Kallosh and A.~Maloney,
arXiv:hep-th/0410076; A.~Sen,
arXiv:hep-th/0411255; D.~Bak, S.~Kim and S.~J.~Rey,
arXiv:hep-th/0501014; A.~Sen,
arXiv:hep-th/0502126;
A.~Dabholkar, F.~Denef, G.~W.~Moore and B.~Pioline,
arXiv:hep-th/0502157; A.~Sen,
arXiv:hep-th/0504005; A.~Sen,
arXiv:hep-th/0505122;
A.~Dabholkar, F.~Denef, G.~W.~Moore and B.~Pioline,
arXiv:hep-th/0507014.

\bibitem{Kraus:2005vz}
  P.~Kraus and F.~Larsen,
  ``Microscopic black hole entropy in theories with higher derivatives,''
  arXiv:hep-th/0506176.

\bibitem{Lin:2004nb}
  H.~Lin, O.~Lunin and J.~Maldacena,
  ``Bubbling AdS space and 1/2 BPS geometries,''
  JHEP {\bf 0410} (2004) 025 [arXiv:hep-th/0409174];

\bibitem{Boni:2005sf} D.~Martelli and J.~F.~Morales, ``Bubbling AdS(3),''
  JHEP {\bf 0502} (2005) 048 [arXiv:hep-th/0412136]; M.~Boni and P.~J.~Silva,
  ``Revisiting the D1/D5 system or bubbling in AdS(3),''
  arXiv:hep-th/0506085.




\end{thebibliography}
\end{document}